\documentclass{aa}
\usepackage{graphicx}
\usepackage{txfonts}
\usepackage{txfonts}
\usepackage{graphics}
\usepackage{epsfig}
\usepackage{natbib}
\bibpunct{(}{)}{;}{a}{}{,}
\begin{document}

%
%
%
%
\newcommand{\rf}{\par\noindent\hangindent 15pt {}}
\newcommand{\etal}{et al.}
\newcommand{\peryear}{yr$^{-1}$}
%
%
%
%
\newcommand{\cmone}{cm$^{-1}$}        
\newcommand{\cmtwo}{cm$^{-2}$}  
\newcommand{\cmthree}{cm$^{-3}$}
\newcommand{\cmq}{cm$^{2}$}
\newcommand{\cmc}{cm$^{3}$}  
\newcommand{\pcone}{pc$^{-1}$}        
\newcommand{\pctwo}{pc$^{-2}$}  
\newcommand{\pcthree}{pc$^{-3}$}
\newcommand{\pcq}{pc$^{2}$}  
\newcommand{\pcc}{pc$^{3}$}  
\newcommand{\kpcone}{kpc$^{-1}$}      
\newcommand{\kpctwo}{kpc$^{-2}$}  
\newcommand{\kpcthree}{kpc$^{-3}$}
\newcommand{\kpcq}{kpc$^{2}$}  
\newcommand{\kpcc}{kpc$^{3}$}  
\newcommand{\kms}{km\,s$^{-1}$}       
\newcommand{\vlsr}{$v_{\rm LSR}$}        
\newcommand{\dv}{$\Delta v$}  
\newcommand{\vs}{$v_{s}$}  
\newcommand{\vsh}{\v$_{shock}$}  
\newcommand{\vr}{v$_{r}$}  
\newcommand{\vrad}{v$_{rad}$}  
\newcommand{\vt}{v$_{t}$}  
\newcommand{\tsys}{T$_{\rm SYS}$}        
\newcommand{\ta}{T$_{\rm A}$}    
\newcommand{\tas}{T$^{*}_{\rm A}$}
\newcommand{\tr}{T$_{\rm R}$}   
\newcommand{\trs}{T$^{*}_{\rm R}$}
\newcommand{\te}{T$_{\rm e}$}  
\newcommand{\tkin}{T$_{\rm kin}$}
\newcommand{\es}{erg s$^{-1}$}                          
\newcommand{\ecs}{erg cm$^{-2}$ s$^{-1}$ }                
\newcommand{\ecssr}{erg cm$^{-2}$ s$^{-1}$ sr$^{-1}$ }
\newcommand{\ecsaa}{erg cm$^{-2}$ s$^{-1}$ \AA$^{-1}$}
\newcommand{\ecsaasr}{erg cm$^{-2}$ s$^{-1}$ \AA$^{-1}$ sr$^{-1}$}
\newcommand{\ecsmu}{erg cm$^{-2}$ s$^{-1}$ $\mu$m$^{-1}$}
\newcommand{\ecsmusr}{erg cm$^{-2}$ s$^{-1}$ $\mu$m$^{-1}$ sr$^{-1}$}
\newcommand{\wc}{W cm$^{-2}$}                            
\newcommand{\wcmu}{W cm$^{-2}$ $\mu$m$^{-1}$}
\newcommand{\wchz}{W cm$^{-2}$ Hz$^{-1}$}
\newcommand{\wm}{W m$^{-2}$}                             
\newcommand{\wmmu}{W m$^{-2}$ $\mu$m$^{-1}$}
\newcommand{\wmhz}{W m$^{-2}$ Hz$^{-1}$}
\newcommand{\um}{$\mu$m}                                 
\newcommand{\molh}{H$_{2}$}                              
\newcommand{\molha}{molecular hydrogen}  
\newcommand{\molhb}{hydrogen molecules}  
\newcommand{\water}{H$_{2}$O}
\newcommand{\watera}{water vapor}  
\newcommand{\waterb}{H$_{2}$O molecules}  
\newcommand{\waterc}{H$_{2}$O maser}  
\newcommand{\lsun}{L$_{\odot}$}                          
\newcommand{\msun}{$M_{\odot}$}
\newcommand{\rsun}{R$_{\odot}$}
\newcommand{\mdot}{\.{M}}
\newcommand{\msunyr}{$M_{\odot}$ $yr^{-1}$} 
\newcommand{\mearth}{$M_{\oplus}$}
\newcommand{\rearth}{$R_{\oplus}$}
\newcommand{\nfss}{$\eta_{\rm fss}$}	
%
%
%
\newcommand{\gapprox}{$\stackrel {>}{_{\sim}}$}   
\newcommand{\lapprox}{$\stackrel {<}{_{\sim}}$}
\newcommand{\about}{$\sim$}                       
\newcommand{\pdown}[1]{\mbox{$_{#1}$}}            
\newcommand{\ppdown}[2]{\mbox{$_{#1_{#2}}$}}      
\newcommand{\pupdown}[2]{\mbox{$^{#1}_{#2}$}}     
\newcommand{\pow}[2]{\mbox{#1$^{#2}$}}            
\newcommand{\powtwo}[1]{2$^{#1}$}
\newcommand{\powsix}[1]{6$^{#1}$}
\newcommand{\powten}[1]{10$^{#1}$}
\newcommand{\iraspsc}{IRAS Point Source Catalogue}
\newcommand{\halpha}{H$\alpha$}                   
\newcommand{\hbeta}{H$\beta$}
\newcommand{\bralpha}{Br$\alpha$}
\newcommand{\brgamma}{Br$\gamma$}
\newcommand{\pfgamma}{Pf$\gamma$}
\newcommand{\sii}{[S\,{\sc ii}] }
\newcommand{\oii}{[O\,{\sc ii}] }
\newcommand{\oiii}{[O\,{\sc iii}] }
\newcommand{\oi}{[O\,{\sc i}] }
\newcommand{\caii}{[Ca\,{\sc ii}] }
\newcommand{\feii}{[Fe\,{\sc ii}] }
\newcommand{\nii}{[N\,{\sc ii}] }
\newcommand{\av}{A$_{V}$}
\newcommand{\magn}{$^{m}$}
\newcommand{\ebv}{E$_{B-V}$}
\newcommand{\amin}{$^{\prime}$}                   
\newcommand{\asec}{$^{\prime \prime}$}
\newcommand{\adeg}{$^{\circ}$}
\newcommand{\afifty}{$\alpha_{1950.0}$}
\newcommand{\dfifty}{$\delta_{1950.0}$}
\newcommand{\rahms}[3]{\mbox{#1$^{\rm h}$#2$^{\rm m}$#3$^{\rm s}$}}
\newcommand{\radot}[4]{\mbox{#1$^{\rm h}$#2$^{\rm m}$#3$\stackrel{\rm s}
{_{\bf\cdot}}$#4}}  
\newcommand{\decdms}[3]{\mbox{#1$^{\circ}$#2$^{\prime}$#3$^{\prime \prime}$}}
\newcommand{\decdot}[4]{\mbox{#1$^{\circ}$ #2$^{\prime}$ #3$\stackrel {\prime 
\prime}{_{\bf \cdot}}$#4}}
\newcommand{\adegdot}[2]{\mbox{#1$\stackrel {\circ}{_{\bf \cdot}}$#2}}
\newcommand{\amindot}[2]{\mbox{#1$\stackrel {\prime}{_{\bf \cdot}}$#2}}
\newcommand{\asecdot}[2]{\mbox{#1$\stackrel {\prime \prime}{_{\bf \cdot}}$#2}}
\newcommand{\ltwo}{$\ell^{\small \rm II}$}
\newcommand{\btwo}{$b^{\small \rm II}$}
\newcommand{\gdeg}[2]{\mbox{#1$\stackrel{\circ}{_{\bf\cdot}}$#2}}  
\newcommand{\elec}{$n_e$}
\newcommand{\jonfrac}{{\it x}}
\newcommand{\ang}{$\AA$}                        
\newcommand{\telec}{T$_e$}			
%
%
%
\newcommand{\CSo}{\mbox{$^{12}$C$^{32}$S}\ }
\newcommand{\CSon}{\mbox{$^{12}$C$^{32}$S}}
\newcommand{\CStf}{\mbox{$^{12}$C$^{34}$S}\ }
\newcommand{\CStfn}{\mbox{$^{12}$C$^{34}$S}}
\newcommand{\CSone}{\mbox{CS (1--0)}\ }
\newcommand{\CSonen}{\mbox{CS (1--0)}}
\newcommand{\CStw}{\mbox{CS (2--1)}\ }
\newcommand{\CStwn}{\mbox{CS (2--1)}}
\newcommand{\CSth}{\mbox{CS (3--2)}\ }
\newcommand{\CSthn}{\mbox{CS (3--2)}}
\newcommand{\CSf}{\mbox{CS (5--4)}\ }
\newcommand{\CSfn}{\mbox{CS (5--4)}}
%
%
%
\newcommand{\COo}{\mbox{$^{12}$CO}\ }
\newcommand{\COon}{\mbox{$^{12}$CO}}
\newcommand{\COt}{\mbox{$^{13}$CO}\ }
\newcommand{\COtn}{\mbox{$^{12}$CO}}
\newcommand{\COe}{\mbox{CO (1--0)}\ }
\newcommand{\COen}{\mbox{CO (1--0)}}
\newcommand{\COz}{\mbox{CO (2--1)}\ }
\newcommand{\COzn}{\mbox{CO (2--1)}}
\newcommand{\COd}{\mbox{CO (3--2)}\ }
\newcommand{\COdn}{\mbox{CO (3--2)}}
\newcommand{\COs}{\mbox{CO (7--6)}\ }
\newcommand{\COsn}{\mbox{CO (7--6)}}
%
%
\newcommand{\abCO}{$^{12}$CO}         
\newcommand{\acCO}{$^{13}$CO}         
\newcommand{\abC}{$^{12}$C}           
\newcommand{\acC}{$^{13}$C}           
\newcommand{\CahB}{C$^{17}$O}         
\newcommand{\CahO}{C$^{18}$O}         
\def\ji{$J=1\rightarrow0$}         
\def\jii{$J=2\rightarrow1$}        
\def\jiii{$J=3\rightarrow2$}       
\def\jiv{$J=4\rightarrow3$}        
\def\jv{$J=5\rightarrow4$}         
\def\jvi{$J=6\rightarrow5$}        
\def\jvii{$J=7\rightarrow6$}       

%
%
%
\newcommand{\hcoplus}{\mbox{HCO$^+$}}
\newcommand{\htcplus}{\mbox{H$^{13}$CO$^+$}}
\newcommand{\hcon}{\mbox{HCO$^{+}$(3--2)}}
\newcommand{\hcn}{\mbox{HCN(3--2)}\ }
\newcommand{\hcnn}{\mbox{HCN(3--2)}}
\newcommand{\hcnc}{\mbox{H$^{13}$CN(3--2)}}
%

%
%

\title{HST and spectroscopic observations of the L1551 IRS5 jets 
(HH154)\thanks{Based on observations made with the Nordic Optical 
Telescope, operated on the island of La Palma jointly by Denmark, Finland, 
Iceland, Norway, and Sweden, in the Spanish Observatorio del Roque de 
los Muchachos of the Instituto de Astrofisica de Canarias, and based
on observations made with the NASA/ESA {\it Hubble Space Telescope},
obtained at the Space Telescope Science Institute, which is operated
by the Association of Universities for Research in Astronomy, Inc.,
under NASA contract NAS5-26555.}}

\author{ C.V.M.~Fridlund\inst{1}\and R.~Liseau\inst{2} \and 
A.A.~Djupvik\inst{3} 
\and M.~Huldtgren\inst{2} \and Glenn.J.~White\inst{2,4} \and F.~Favata\inst{1} 
\and G.~Giardino\inst{1}}

\offprints{ C.V.M. Fridlund -- email: Malcolm.Fridlund@esa.int}

\institute{ESA Astrophysics Missions Division, Research and Scientific 
Support Department, 
ESTEC, P.O. Box 299, NL\,--\,2200~AG Noordwijk, The Netherlands
\and Stockholm Observatory, SCFAB, Roslagstullsbacken 21, S-106 91 
Stockholm, Sweden
\and Nordic Optical Telescope, Apartado 474 E-38700 Santa Cruz de La Palma 
Canarias, Spain
\and Center for Astrophysics and Planetary Science, University 
of Kent, Canterbury, Kent, CT2 7NZ, UK}
   
\date{Received ; accepted }

\abstract{
 We have carried out a thorough optical study of the closest star 
formation jet. The inner arcminute surrounding the class 0/I binary 
protostar L1551 IRS5 and its associated jet (HH154) have been observed using 
the Hubble Space Telescope with the WFPC2 camera, the ESO New 
Technology Telescope with the  EMMI spectrograph and 
the Nordic Optical Telescope with the  ALFOSC spectrograph. 
This data set is compared to earlier ground based imaging 
with the aim to study the evolution of this particular jet, and its 
possible interaction with the molecular material in the bipolar 
molecular outflow associated with this source. The velocity field of 
the jet is mapped out. The highest velocities are found in the 
vicinity of the recently discovered X-ray source emanating from a shock in 
this jet. The energy radiated by the X-ray source is compatible with 
these velocities. The \halpha~and \hbeta~emission from the jet is used 
to determine the extinction, which is found to increase inwards in the 
jet towards the protostar. The extinction towards the X-ray source is 
consistent with the one determined from the X-ray spectrum.
}
\maketitle
%

\section{Introduction}
Among bipolar molecular outflows, the one centered on the young stellar 
object IRS5 in L1551 was the first discovered \citep{sne80}. It is 
relatively nearby, \about 150pc \citep{kdh94,re94}, allowing 
the study of small 
spatial elements, and oriented with its major outflow axis most 
likely at an angle of between 45\adeg~and 60\adeg~\citep{lf05}~with 
respect to the plane of the sky. This results in 
red- and blue-shifted CO outflow lobes which are spatially well 
separated, and which have an angular extent of several tens of arcminutes 
on the sky. The red- and the blue-shifted CO lobes contains several 
Herbig Haro objects (HH objects) most prominently seen in the 
blue-shifted lobe, (e.g. HH28, HH29). In \cite{ch79} 
and \cite{fsnd84,flp93,flg98} the authors argue, based on proper 
motion vectors, velocity gradients in CO, and the radial velocity 
field, that the outflow is originating at IRS5 and is interacting with the 
ambient medium at the postion of the 
Herbig Haro objects. In \cite{dev1999,dev2000}, the authors on the other 
hand argue that the HH28 and HH29 are excited by a flow from another YSO, i.e. 
L1551NE.

 The low mass characteristics of IRS5 were determined by observations in the 
far infrared \citep{fnd+80}. A 
jet has been observed at optical wavelengths \citep{mun83,nec87,fl94,fl98}, 
in the near IR \citep{cam88,pyo02,pyo05}~and in the radio 
continuum \citep{cbs82,rod86}. The realisation that this jet has 
Herbig Haro characteristics \citep{sne85} has led to it receiving the  
designation HH154 in the compilation of \citet{re94}. Morphological 
changes were detected within the jet by \cite{sne85} and \cite{nec87}. The 
latter found that 
over the epoch $1983 - 1987$, a new knot appeared near IRS5, and that 
the transverse velocities for several knots 
in the jet were of order \about~190 \kms\ -- in agreement with earlier proper 
motion measurements of one of the knots by \citet{sne85}. 
\cite{nec87}~introduced a consistent 
nomenclature for the features observed along 
the jet. This was also used by \cite{fl94}~-- hereafter FL94 and \cite{fl98} 
-- hereafter FL98, and we 
continue to use their designation in this paper, consistently updated 
in view of the evolution of the jet. 

 At all observed wavelengths, the jet is relatively compact compared to the 
molecular outflow, growing from \about~7 arcsec to \about 14 arcsec 
between 1983 and 2001. The jet 
terminates in a bright emission knot -- `D' according to the 
nomenclature described above -- although 
some fainter, isolated emission peaks (A, B and C) are also detected 
further `downstream'. \citet{mun91} used image restoration techniques 
on a deep \sii~image and suggested that the jet consists of a limb 
brightened cavity, with a high velocity wind permeating the cavity, 
which through interactions with the walls of the cavity, produce the HH 
knots. The optical properties of the jet were then further elucidated by 
FL94 who reported on high spatial resolution monitoring observations of the 
L1551 IRS5 jet carried out between 1989 and 
1993, made from the ground in excellent seeing conditions. FL94 confirmed 
the earlier results by \cite{nec87}
that morphological changes along the jet have dynamical time-scales 
of $<$~a few years, 
and that these changes can be interpreted as proper motions of knots within 
the jet. Disappearance of `old' and appearance of `new' knots was recorded. 
It was also found that the brightest feature (D) in the jet accounts for more 
than 
50\% of the observed total emission in the lines of \halpha\ and 
\sii6716,6731\AA. Also from these observations, three 
different transversal velocity components (120 \kms, 190 \kms~and $> 310$ \kms) 
were identified. Taken together with earlier radial velocity measurements 
found in the literature, FL94 hypothesized that the spatial 
distribution of the knots was best represented by the features moving close 
to the surface of a conical volume. FL98, which reports preliminary results 
from the first visit in the two-epoch Hubble Space Telescope (HST) program 
fully described in the current paper, interprets their data together with some 
spectroscopic data as indicative of the existence of two separate jet-flows 
originating from IRS5. Each of these is possibly originating from one of the 
two components of IRS5 found 
by \citet{loo97}. FL98 also identified the brightest component (`D') of the 
jet with the `working surface' of the `northern jet' and discussed the 
structure of that shock. Using the models of \citet{har89}, they found that 
the ionisation 
fraction appears high (\about~50\%~-- 100\%). From the velocities 
determined, an estimate of the forces required to accelerate the individual 
knots could be made. This was compared to the force required to accelerate 
the molecular gas {\em in the same volume of space as the jet} 
(Fridlund \& Knee, 1993) and 
found to be at least two orders of magnitude too low assuming an ionisation 
fraction of \about~unity. FL98 pointed out the impact of this 
result on the question of the level of ionisation in optical jets 
which has turned into a major problem when trying to 
unify models for optically visible jets with those of the accompanying 
molecular outflows.

 The structures closest to IRS5 have been observed with 
interferometers in the radio continuum (1.4 GHz to \about 220 GHz), by a 
number of researchers \citep{bc85,setal85,rod86,km90,lfp+95,loo97}. The 
latter authors pointed out that a 3 component model emerges with a large scale 
(10~to 30 arcsec) envelope, a disk like feature (\about~1 arcsec) and a 
binary source with individual components separated by 0.35 arcsec, and where 
the disks have each a FWHM of \lapprox~0.3 arcsec. This double 
source, which was originally discovered by \cite{bc85}~and at that time 
interpreted by \cite{rod86}~as the 
inner ionized edges of a gas and dust toroid, is now seen as 
a young binary system \citep [also suggested by][]{bc85}~with a 
separation of \about~50AU \citep{loo97}. \citep{rod03a,rod03b} have 
analysed 
interferometric radio continuum observations, and trace the two jets 
to their origin. Based on that data, those authors suggest that the 
northern jet-component emanates from the southern binary and vice versa. 
It has also 
been demonstrated \citep{fbwpt02,white2000}~that there 
is also a fourth component in the form of a large 
(\about~15\,000 AU), massive (several \msun), rotating flattened envelope 
of molecular gas. 
 
 In the L1551 cloud, we find the astrophysical jets located nearest to 
the Earth (HH154, HH30 and the HL Tau jet). This implies that for 
HH154, the HST can achieve similar (\about~10$^{14}$~cm), or better, spatial 
resolution 
to that used in current 2D hydro-dynamical calculations of 
jets \citep{rag04}.  A multi-cycle 
observing programme with the HST consequently appeared well motivated, with the 
primary goal being to study the time evolution of the jet, to use 
the shock diagnostic emission lines of \halpha\ and 
the \sii~doublet at $\lambda \lambda 6717,6731$\AA~in order to 
determine its physical properties. To detect fainter details we also 
obtained $R$-band (F675W filter) observations. Further we wished to study the 
structure furthest in towards IRS5. Since the jet here suddenly appears from 
behind \about~100 magnitudes of obscuring material (presumably the 
accretion disk), we try to observe light scattered 
up along the major axis of the jet. To achieve this we selected the 
F814W I-band filter. As mentioned above, preliminary results have 
been reported in \citet{fhl97}~and FL98. Included 
in the latter paper are some (also preliminary) ground based 
spectroscopic data, demonstrating a radial velocity gradient along the 
jet, as well as two separate velocity fields. 

 Recently, X-rays emanating from the jet have been discovered \citep{fav02}. 
The observations carried out with XMM-Newton have a positional uncertainty of 
roughly the same size as 
the jet. The X-ray source with a temperature of a few million degrees 
and a visual extinction of \about~5 magnitudes, was tentatively 
identified as originating from the working surface ('knot D'), mainly 
because of its brightness, and feature D being the only source with 
reported \oiii~emission then known \citep{cf85,sne85}. The X-rays had to 
originate in the jet, since the star itself is hidden by the very 
large extinction from the surrounding disk.
The position of the X-ray source have recently been measured by 
\citet{fei03}, who observed with the Chandra X-ray observatory 
providing a spatial resolution of \about~1 arcsec. This has enabled them 
to determine that the X-rays emanate roughly from \about~0.5 arcsec to 1 
arcsec from the radio position of IRS5. 

 In the present paper we give the detailed results and interpretation from 
two HST epochs (cycle 5 \& cycle 6). We also use new, extensive 
ground based imaging and spectroscopic data to support our interpretation. 

 We confirm the dual structure of the jet itself, and we describe the 
somewhat surprising morphology found close to the 
base of the jet. The time evolution of the jet is described, as 
discerned from the measured proper motions of individual knots within 
the jet. We also discuss the structure of different individual shocks, 
including the excitation conditions and the electron density along the 
jet and within separate knots, identify the velocity component 
associated with the X-ray source and the implications of having 
a strong X-ray source illuminating the accretion disk surrounding IRS5.
\section{Observations and data reduction}
\subsection{HST Imaging}

 The area surrounding the L1551 IRS5 jet was imaged using the Wide 
Field and Planetary Camera 2 (WFPC2 -- Trauger \etal, 1994) 
aboard the Hubble Space Telescope. This 
camera consists of 4 adjacent 800 $\times$~800 Loral CCD's. Three of 
these arrays have an image scale of 0.1 arcsec~${\rm pix}^{-1}$~(WF-chips), 
while the fourth -- the planetary camera (PC-chip) -- has an image scale of 
0.046 arcsec~${\rm pix}^{-1}$, thus well sampling the diffraction pattern 
in the $R$- and $I$-band. The FOV in the planetary camera is 
25\asec~$\times$~25\asec and since the jet visible in the ground 
imagery is less than 15 arcsec 
long, we centered the PC on the position of IRS5. We used an orientation 
that allowed us to detect the only visible star within the field-of-view on one 
of the WF chips, leading to potential relative astrometry between 
epochs of \about~0.1 arcsec. With this orientation we unfortunately 
could not image any of the other brighter nebulosities associated with L1551 
IRS5. Note that the brightest object, HH29, has been imaged 
by \citet{dev2000}. Some fainter 
nebulosities are, however, visible on one of the WF chips. 
 In this paper we only discuss features relevant to the jet. 

 We observed HH154 during 3 HST visits. During each visit, two exposures each 
were obtained through the F656N (H$\alpha$) and the 
F673N ($\lambda$6717\AA+6731\AA) filters beginning on 1996.034. Total 
exposure time for each filter was 2\,500 s. On 1996.078 we obtained 3 exposures 
each through the F675W ($R$) and F814W ($I$) filters with the total 
observing times being 1\,800 s and 2\,400 s respectively. We again obtained data 
through the F656N (H$\alpha$) and F673N ($\lambda$6717\AA+6731\AA) filters on 
1998.074, providing us with a time base of almost exactly two years 
(727 days).

 The data were reduced and calibrated into units 
of \ecs~using standard 
methods as described in e.g. \citet{hol95}, \citet{heat96}, 
\citet{rei97}~and the WFPC2 handbook. 

 All the images were registered with each other by using the stellar image of 
VSS4 and the rotation angle information. The accuracy in this process 
is estimated to be \about~0.1 arcsec. 
%
%
\begin{table}[htb]
\begin{center}
\caption{Observed fluxes (\ecs) for the assumed working surface (knot D), the 
feature F, and the total jet in  in all of the imaging data. Note that 
these data have not been de-reddened. The quoted accuracy includes both the 
estimated precision in the flux calibration {\em and}~the measurement error.}
\label{mytable}
\begin{tabular}{llllll}
\hline
Feature& H$\alpha$& H$\beta$&\sii& Source & Accu-\\
& & & & & racy (\%) \\
\hline
D & 1.8~$\times~10^{-14}$ &--& 1.8~$\times~10^{-14}$ & FL94 & 10\\
F & 5.2~$\times~10^{-15}$ &--&  3.0~$\times~10^{-15}$  & FL94 & 10 \\
$\Sigma$$_{All jet}$& 3.3~$\times~10^{-14}$  &--& 2.9~$\times~10^{-14}$  & FL94 & 10 \\
D	& 2.1~$\times~10^{-14}$  &--& 2.2~$\times~10^{-14}$  & HST 1996 & 5 \\
F       &   3.8~$\times~10^{-15}$  &--& 3.4~$\times~10^{-15}$  & HST 1996 & 5 \\
$\Sigma$$_{All jet}$& 5.2~$\times~10^{-14}$  &--& 4.8~$\times~10^{-14}$  & HST 1996 & 5 \\
D & 1.8~$\times~10^{-14}$ & 2.3~$\times~10^{-15}$  & --& NOT 1996 & 10 \\
F & 2.7~$\times~10^{-15}$  & 2.3~$\times~10^{-16}$  & --& NOT 1996 & 10 \\
$\Sigma$$_{All jet}$& 4.1~$\times~10^{-14}$  & 5.0~$\times~10^{-15}$  & --& NOT 1996 & 10 \\
D	& 1.8~$\times~10^{-14}$ &--& 2.0~$\times~10^{-14}$ & HST 1998 & 5 \\
F       &   3.4~$\times~10^{-15}$  &--&   3.3~$\times~10^{-15}$    & HST 1998 & 5 \\
$\Sigma$$_{All jet}$ & 4.3~$\times~10^{-14}$  &--& 4.2~$\times~10^{-14}$ & HST 1998 & 5 \\
D & 2.0~$\times~10^{-14}$  & --& 1.7~$\times~10^{-14}$  & NTT 1998 & 10 \\
D & 1.8~$\times~10^{-14}$ & 2.4~$\times~10^{-15}$  & --& NOT 2001 & 10 \\
F & 3.8~$\times~10^{-15}$  & 3.2~$\times~10^{-16}$  & --& NOT 2001 & 10 \\
$\Sigma$$_{All jet}$& 4.1~$\times~10^{-14}$  & 5.2~$\times~10^{-15}$  & --& NOT 2001 & 10  \\
\hline
\end{tabular}
\end{center}
\end{table}
%
%
%
%
%
\begin{table}[htb]
\begin{center}
\caption{Extinction determined from the \halpha/\hbeta~ratio, at 
two epochs (December 1996 and October 2001) for several features in 
the jet (see text). The accuracy in the determination of the ratios is 
\about~20\%. Data taken from \citet{im05} }
\label{exttable}
\begin{tabular}{lllll}
\hline
Feature& Ratio$_{1996}$ & Ratio$_{2001}$ & A$_V$(1996) & A$_V$(2001) 
\\
& & & {\em R = 3} &{\em R = 3}\\
\hline
D & 8.5& 7.0& 2.8& 2.2 \\
E &  8.0 & 10. & 2.6 & 3.2 \\
F & 11.9 & 11.9 & 3.7 & 3.7 \\
Region between & 12 -- 15 & 14 -- 20 & 4 -- 4.5 & 4.2 
-- 5.1 \\
F and end & & & & \\
of jet (F2) & & & & \\
\hline
\end{tabular}
\end{center}
\end{table}
\subsection{Ground based imaging}
 In December 1996 and in October 2001, calibrated images were obtained 
using the Nordic Optical Telescope. The instrument combination was 
identical to that used in the ground based spectroscopy except that 
ALFOSC was set to imaging mode. Images were obtained through a number 
of filters, of which results for \halpha, \hbeta, \sii4067,68\AA~and 
\oiii5007\AA~are reported here, while a more detailed description, as 
well as the presentation of further data is deferred to a later paper.
\subsection{High resolution spectroscopy}
 Spectroscopic observations of the jet were carried out with 
the NTT 3.58m telescope -- EMMI (ESO Multi-Mode 
Instrument) combination at the European Southern 
Observatory, La Silla, Chile. The observation was carried out on the 
night of 22 December 1998. We used the echelle mode with grating No. 10 
and grism No 6, resulting in a wavelength coverage of 6\,000\AA~-- 8\,400 
\AA. Using a 1\asec $\times$ 10\asec~slit, we have a spectral resolution of 
28\,000 (equivalent to a velocity resolution of \about~11~\kms~at \halpha). 
The seeing was about 1 arcsec, and three slit positions, each separated by 
the same amount in declination,  were observed. These observations only cover 
the working surface of the jet (feature D). Lines detected with the 
measured fluxes and velocities are listed in Table \ref{ion1}. 
The reduction of the data were carried out with the ECHELLE context 
within the ESO MIDAS reduction package, and with flux calibration 
being obtained by observing the flux standard HR1544 
(see http://www.hq.eso.org/observing/standards/spectra for details).
%
%
\begin{table*}[htb]
\begin{center}
\caption{Ions detected in spectroscopic observations with the ESO 
NTT-EMMI instrument. These observations only covered knot D 
(brightest feature in Figure \ref{mypetjet}). Note that the data are {\em 
not}~de-reddened.
Note also that the values in the table refer to the {\em maximum} 
values. There is a large variation between slit positions for 
different lines. These differences presumably are due to a geometric effect 
(working surface spatially resolved), discussed in the text. The 
accuracy in the flux determination is approximately 10\%, and this 
error includes both the error in the calibration and the actual 
measurement error. The error in velocity is 10 \kms.}
\label{ion1}
\begin{flushleft}
\begin{tabular}{llllll}
\hline
Ion&Rest wavelength & Max line width  & Max peak velocity & Total flux 
observed& Comment\\
&(\AA)&(FWZI \kms) & (\kms) & (10$^{-15}$~\ecs)& \\
\hline
{\rm [OI]}&6300.2&180&$-170$&5.2&\\
{\rm [OI]}&6363.8& 185& $-175$& 5.3&\\
{\rm [NII]}&6548.1& 65&$-185$&0.63 &\\
\halpha&6562.8&310&$-175$&20.0&\\
{\rm [NII]}&6583.6& 270& $-185$& 7.0& several separate components$^1$ \\
{\rm [SII]}&6716.4&200& $-175$& 7.26& 
$\Sigma_{flux(6717+6731)}~=~17.2~\times~10^{-15}$~\ecs \\
{\rm [SII]}&6730.8& 220&$-170$&9.9&\\
{\rm [FeII]}&7155.1& 250& $-170$&2.6&\\
{\rm [CaII]}&7291.5& 240&$-180$&6.5& several separate components$^1$ \\
{\rm [OII]}&7319.4& -- &-- & -- & \\
{\rm [CaII]}&7323.9& 125& $-160$ & 3.0 & \\
{\rm [OII]}&7329.9&  -- & -- & 0.16 & Detected at 3$\sigma$-level\\
\hline
\end{tabular}
$^1$~Values in table refer to the strongest component
\end{flushleft}
\end{center}
\end{table*}
%
%
%
%
%
\begin{table*}[htb]
\begin{center}
\caption{Ions detected in medium resolution spectroscopic observations 
with the Nordic Optical Telescope ALFOSC instrument. These observations 
covered essentially the whole jet (one slit position 
missing, see Figure \ref{alfoscspectra}).
Note also that the values in the table refer to the {\em maximum} 
values. There is a large variation between different slit positions 
for {\em the same} ions. These differences presumably are caused by 
the two {\em distinct} jets. The total flux refers to the flux from 
{\em all} of the jet (excluding that emanating from the missing slit 
position). Note that the data is {\em not}~de-reddened. The accuracy 
in determining the flux is 10\%, which includes both calibration error 
and the actual measurement uncertainty. The velocities can be 
estimated to a precision of \about~1/2~of a pixel, corresponding to 
about 25 \kms.}
\label{ion2}
\begin{flushleft}
\begin{tabular}{llllllll}
\hline
Ion&$\lambda_0$ & Feature &  FWZI& Peak Vel. & F$_{tot}$ & Max vel. & Comment\\
&(\AA)& & (\kms) & (\kms) & (10$^{-15}$~\ecs)& (\kms) & \\
\hline
[NII] &6548.1& D& 200& $-$160& 2.5& $-260$& \\
&&E&-&$-190$&0.34&$-$&\\
&&F&200&$-210$&0.72&$-305$&\\
&&F2&300&$-$&0.49&$-400$&\\
\hline
$H\alpha$&6562.817&D&300&$-150$&17.23&$-310$& \\
&&E&270&$-128$&1.43&$-243$&\\
&&F&190&$-220$&4.1&$-310$&\\
&&F2&160&$-243$&0.82&$-310$&\\
&&F3&100&$-50$&0.72&$-105$&\\
\hline
[NII]&6583.6& D &300 &$-160$ &6.2 &$-320$ & \\
& & E&-&-&-&-& \\
& & F& 275& $-250$& 2.7& $-345$ & \\
& & F2& 250& $-250$& 0.67& $-410$& \\
\hline
[SII]&6716.42& D& 275&$-215$ &5.4 &$-325$ & $\Sigma_{flux(6717+6731)}~= 
12.2^a$\\
& & E&-&-&-&-& \\
& & F& 180& -280& 1.35 & -395& $\Sigma_{flux(6717+6731)}~= 3.0^a $\\
& & F2& 400& -& 0.5 & -360&$\Sigma_{flux(6717+6731)}~= 1.43^a $\\
& & F3& 180& -125 & 1.49 & -215&$\Sigma_{flux(6717+6731)}~= 3.6^a $\\
\hline
[SII]&6730.78&D & 275 & -160 & 6.9 & -325 & \\
& & E & 180 & -160 & 0.54 & -250 & \\
& & F & 180 & -260  & 1.65 & -395 & \\
& & F2 & 400 & -70 & 0.93 & -360 & \\
& & F3 & 180 & -115 & 2.1 & -215 & \\
\hline
[FeII]&7155.14& D & 210& -170& 2.4 & -270& FWHM = 160 \kms$^b$\\
& & F & 300 & -295 & 2.6 & -420 & FWHM = 140 \kms \\
& & F2 & 250 & -330 & 0.78 & -430 & FWHM = 150 \kms \\
\hline
[FeII]&7172.& D & -- & --& 0.33& -- & \\
\hline
[CaII]&7291.46& D & 250 & -165 & 6.1 & -290 & \\
\hline
[NiII]&7379.6& D & - & - & 1.95 & - & $^b$ \\
\hline
[FeII]& 7452.54 & D & - & - & 0.86 & - & $^b$ \\
\hline
[FeII]&8619.33& D & 250 & -120 & 0.64 & -250& $^b$ \\
\hline
$^a$ \ecs \\
$^b$Within a 2\asec~$x$ 2\asec~aperture \\
\end{tabular}
\end{flushleft}
\end{center}
\end{table*}
\subsection{Medium resolution spectroscopy}

 The second set of spectroscopic observations was obtained at the 
Nordic Optical Telescope (2.6m) in La Palma on the nights of 28 \& 29 December 
1999. We used the ALFOSC imager/spectrograph. This instrument was 
used in echelle mode and was 
equipped with a Loral 2k$\times$2k CCD with 15 $\mu$m pixels 
corresponding to 0.189 arcsec~pixel$^{-1}$. The slit was 13.8 arcsec~long, 
but due to crowding of the higher orders, only 12 arcsec has been used. The 
slit width utilized was 1 arcsec. The 
grating/grism combination has a practical wavelength coverage of 
4\,800\AA~-- 11\,000\AA, with a spectral resolution of 4\,300 resulting in a 
velocity resolution of \about~70 \kms~at \halpha. Special care was taken in 
achieving precision pointing of the slit on the sky. Thus a short 
(20 s) exposure through an $R$-filter was obtained at each position. Then 
without moving the telescope, an image of the slit was made on the 
detector. Finally, the spectrum was recorded followed by another slit 
image and $R$-band image before moving to the next postion. This allows 
us to spatially register our spectra to within one pixel 
(0.189 arcsec). A total of four (4) positions was recorded, with 
an integration time of 3\,600 seconds per position. The positions of the 
slits on the sky can be 
seen in Figure \ref{alfoscspectra}. Also here the reductions of the data 
were carried out with the ECHELLE context within the ESO MIDAS reduction 
package, and with flux calibration 
being obtained by observing the two flux standards hz4 and 
BD+28\adeg4211 (see http://www.hq.eso.org/observing/standards/spectra 
for details).
\section {Results}
\subsection{The wide band HST data}

\begin{figure*}
 \begin{center}
    \leavevmode
\includegraphics[totalheight=15cm,angle=-90]{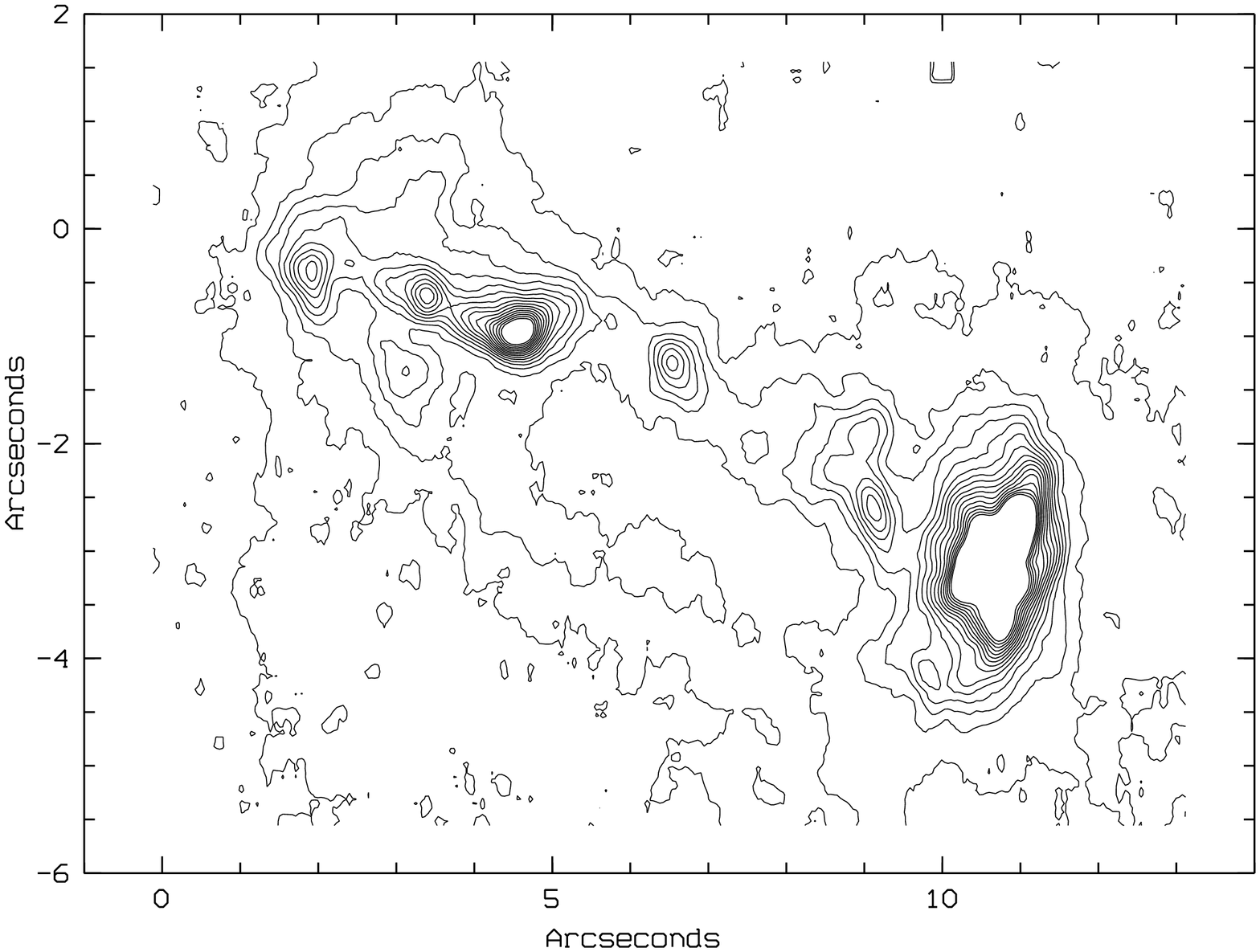}
\includegraphics[totalheight=15cm,angle=-90]{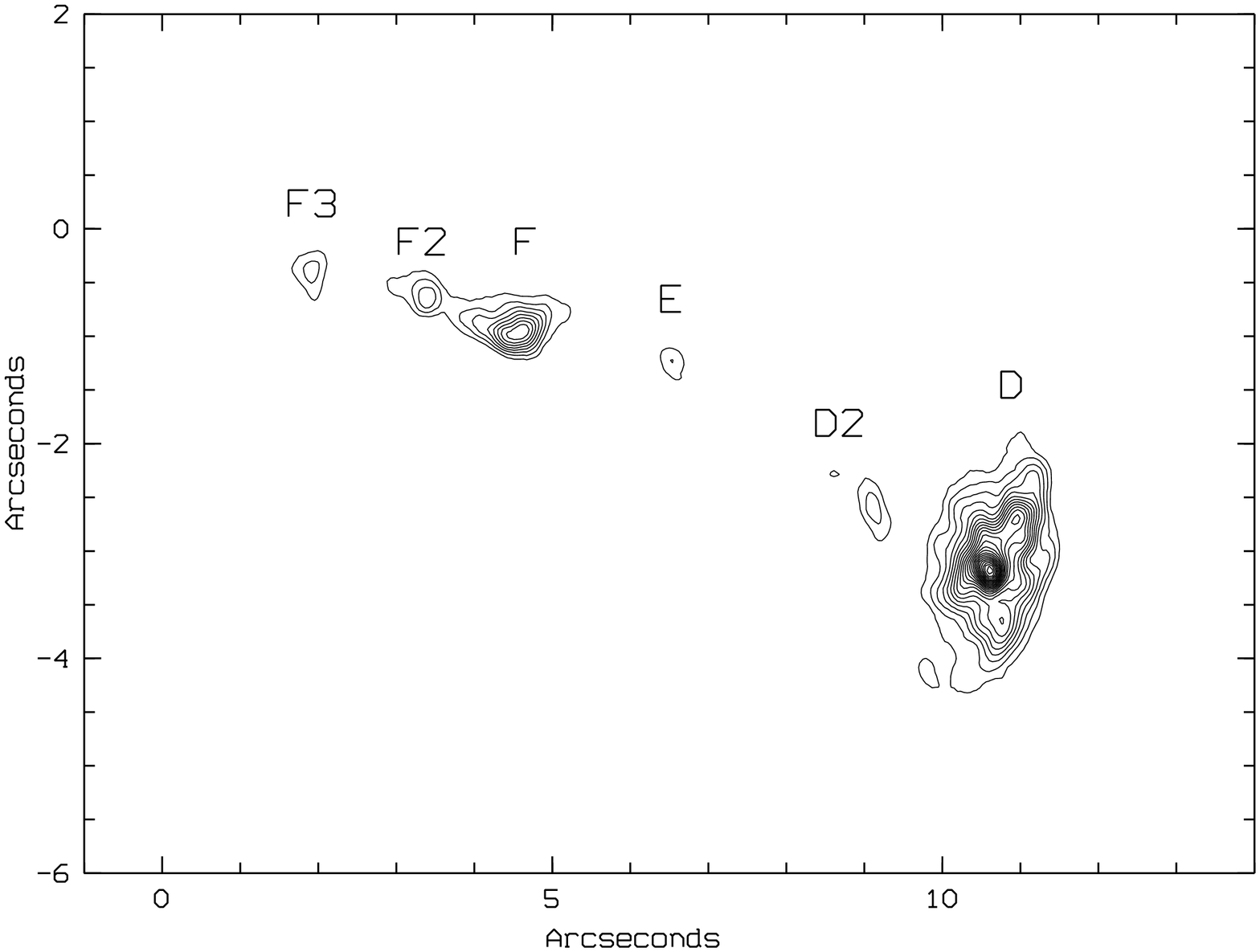}
    \caption{The 1996 HST R-band image displayed as contour plots at 
two different 
levels of contrast in order to show both the faint structures and the details 
in the feature F -- D (see text). The 0\asec,0\asec position is the nominal 
L1551 IRS5 position. In the top panel, the lowest contour 
is $2.5\times10^{-14}$~\ecs, and each successive contour is separated 
by  $4.0\times10^{-15}$~\ecs. In the bottom panel, the values are 
$5.0\times10^{-14}$~\ecs, and $8.0\times10^{-15}$~\ecs~respectively.}
\label{mypetjet}
  \end{center}
\end{figure*}

\begin{figure*}
 \begin{center}
    \leavevmode
    \epsfig{file=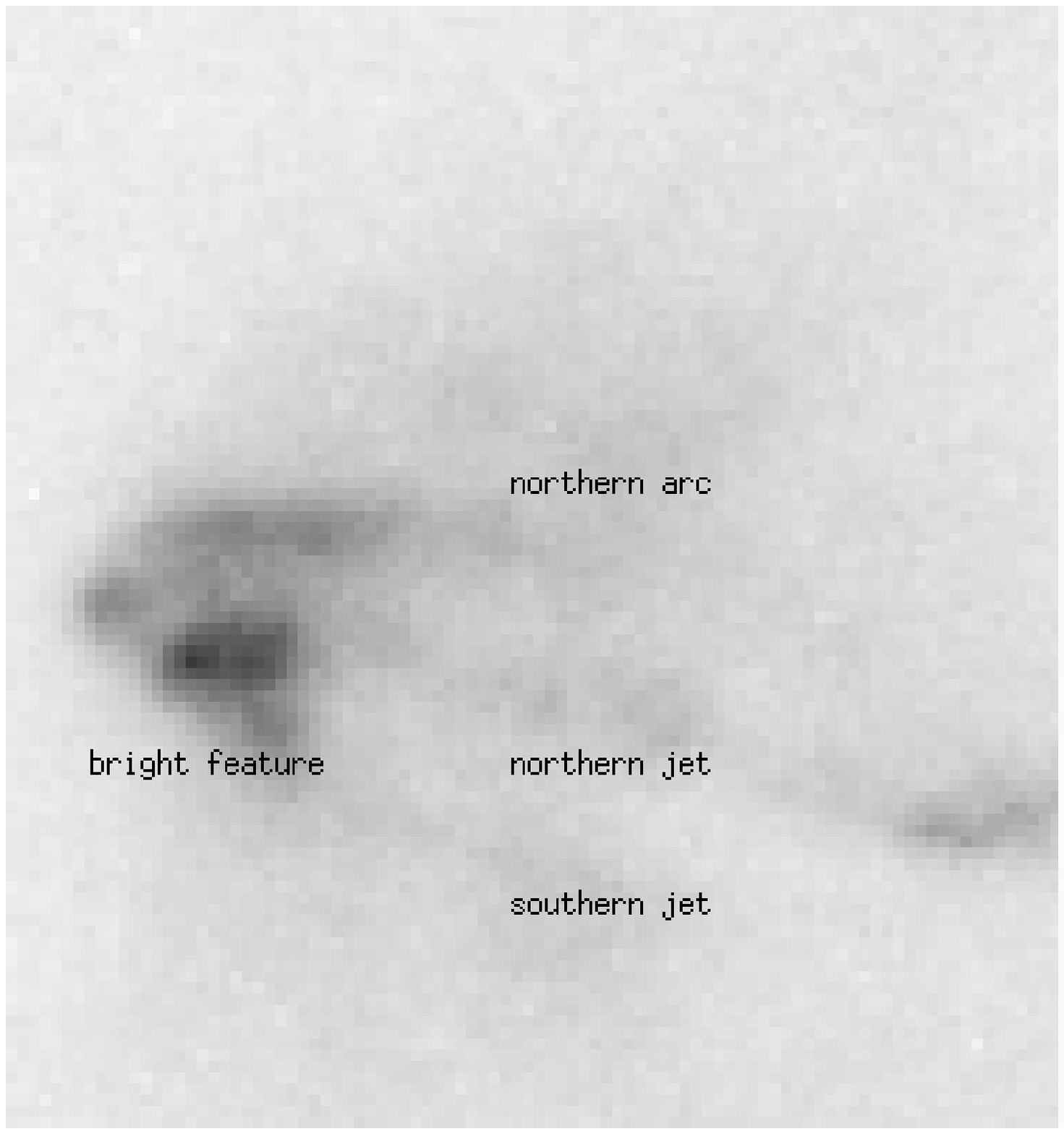, width=10cm, bbllx=70pt,
      bblly=250pt, bburx=522pt, bbury=755pt, clip=}
\includegraphics[totalheight=15cm,angle=-90]{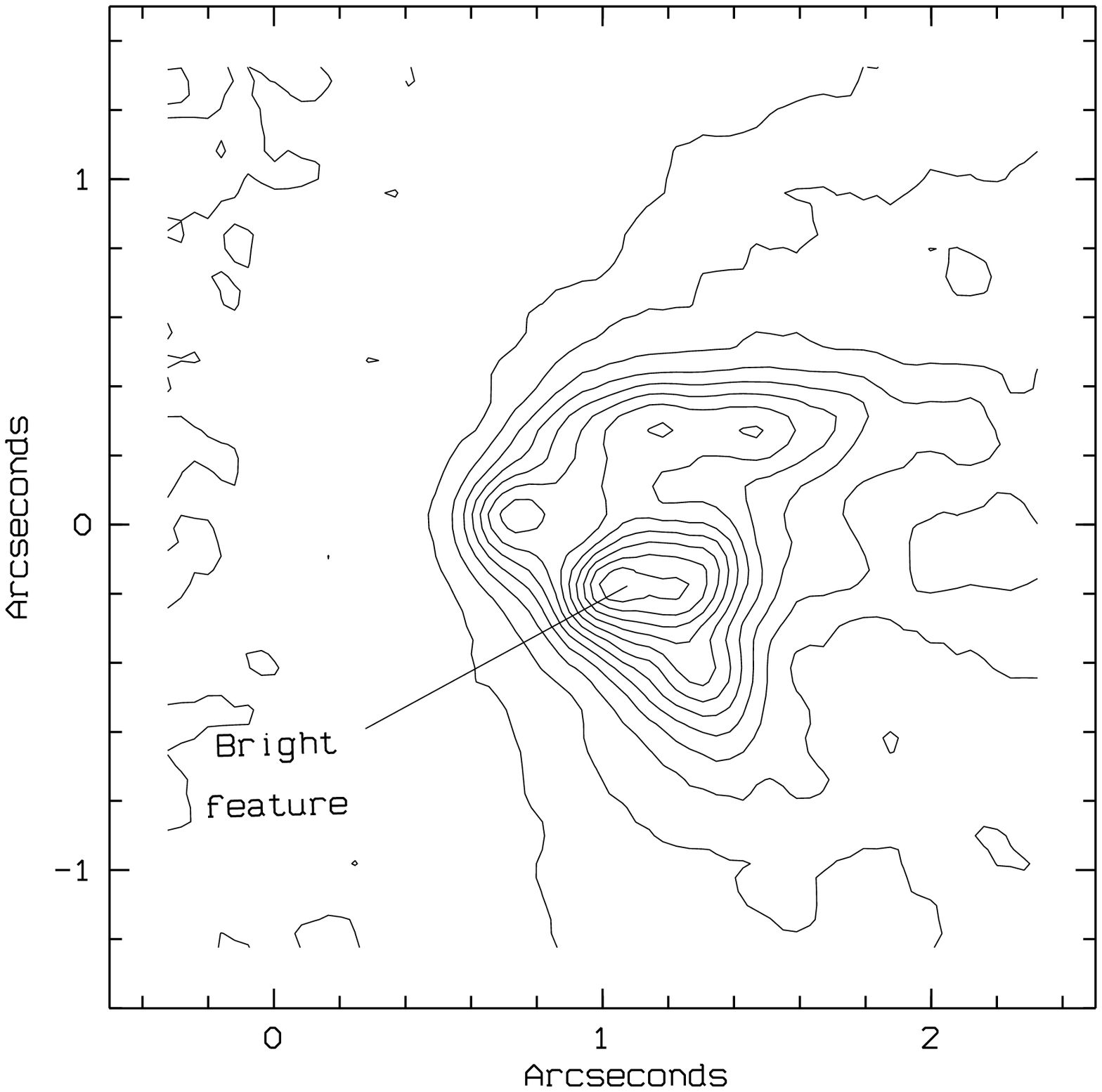}
    \caption{The inner part of the 1996 HST I-band image displayed as 
both a contour plot and a grey scale in order to show both faint structures 
and the details in the "bright feature (see text). The 0\asec,0\asec 
position is the nominal L1551 IRS5 position. In the bottom panel, the 
lowest contour value is $5.0\times10^{-17}$~\ecs, and each contour is 
separated by $1.4\times10^{-17}$~\ecs. This data has been smoothed by 
a Gaussian filter with a 2 pixel radius.}
\label{wowjet}
  \end{center}
\end{figure*}

 In Figure \ref{mypetjet}, we show the $R$-band representation of the 
whole jet including the location of the 
clumps F -- D, using designations originally introduced by 
\citet{nec87}, and also used and modified by FL94 and FL98. 
Then, in Figure \ref{wowjet} we present part of the $I$-band images of the 
jet. The raw images have a pixel resolution of 
0.046 arcsec per pixel. The HST is diffraction limited at the 
wavelengths in question in this paper, and the PC chip is therefore 
Nyqvist sampling the PSF. The image scale corresponds at the assumed 
distance of L1551 to \about~7 AU or 10$^{14}$~cm per pixel. The data 
are presented 
at two different levels of contrast in order to bring forth both the bright 
small scale structure and the faint extended emission tracing the outline 
of the jet. The $R$- and $I$-band data both show a 
highly structured jet, but with significant differences. In 
the R-band we can 
discern several major arc-like components, each consisting of a 
number of knotlike features. Most obvious, a relatively 
bright, narrow and `patchy' structure with a clear `wiggle' ends in a 
very bright feature \about~10 -- 12 arcsec downstream. To conform to FL98 we will 
designate this the northern jet. The second component is fainter, has only 
1 or 2 knot-like features before bending back and intersecting the 
bright feature at the end of the northern jet. Here we designate this 
feature the southern jet. We also see a faint very narrow and short arc 
(designated the northern arc), a few tenths of an arcsec north of the 
bright northern jet (see also Figure \ref{wowjet}).

 The $I$-band flux in 
the inner regions of the jet, are enhanced relative to the flux in the 
$R$-band image, while the northern jet is relatively weaker in the 
$I$-band images. All the knots and features visible in the $R$-band can, 
however, be detected also in the $I$-frame. The $R$-band filter admits 
emission from the \oi~6\,300, 6\,364\AA, \nii~6\,548, 6\,583\AA, \halpha\ and 
the \sii~6\,717, 6\,731\AA~lines, as well as some scattered light.  

 A few emission lines fall within the band-pass of the $I$-filter 
{\it viz.}~\feii~7\,155\AA, \caii~7\,291\AA~and \oii~+ 
\caii~7\,325\AA~-- see also \cite{cf85}. Although, these lines can be seen in 
our medium and high resolution spectra of the jet to be at significant flux 
levels (\about~$1\over2$~F(\halpha)~for feature D) for a few of the features, 
as what concerns the innermost part of the jet, 
they are essentially nonexistent -- see Table \ref{ion2}. The spatial 
distribution of this emission -- as seen in the spectra -- is essentially 
identical to that of the stronger emission lines (e.g. \sii~-- see 
Figure \ref{alfoscspectra}). Features detected in the $I$-band image that have 
no or little counterpart in the \halpha~and \sii~images we therefore judge to 
be primarily due to scattered light.
\subsection{Narrow band HST data}
\begin{figure*}
 \begin{center}
    \leavevmode
\includegraphics[totalheight=20cm,angle=-90]{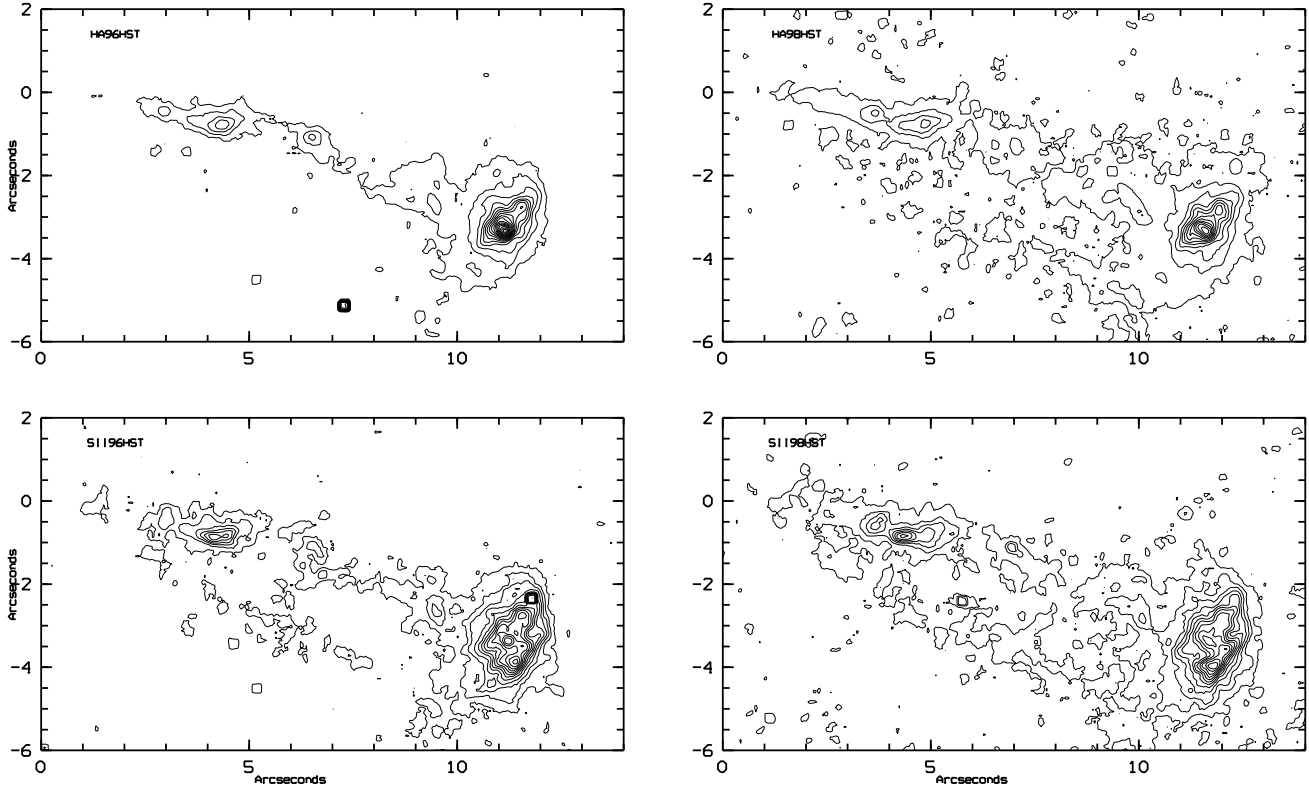}
    \caption{The 1996 (left upper panel) and 1998 (right upper 
panel) \halpha~images displayed as contour plots. The lowest contour 
is at the 3$\sigma$ level, and each consequtive contour is at intervals of 
1$\sigma$. The 0\asec,0\asec position is the same as in Figures 1 and 
2. Note that the innermost part of the jet (position F3 in the R-band 
image) is not seen in emission. The bright spike in the 
top panel (\asecdot{7}{3},-\asecdot{4}{8}~is a Cosmic ray hit.) The 
1996 \sii~image (bottom left panel) and 1998 \sii~(bottom right panel) 
imagesare also displayed as contour plots with the same contours. 
The bright pointlike knot in the lower left panel at 
position \asecdot{11}{5},-\asecdot{2}{5} is a cosmic ray strike that could 
not be removed.
All images are obtained with the same exposure time.} 
\label{siicontour}
  \end{center}
\end{figure*}

In Figure \ref{siicontour} we display contour 
plots of our \halpha~and \sii~chosen to both bring out faint details 
and show changes due to e.g. proper motions.

 No trace of the `northern arc', so clearly visible in 
the I - frame, can be traced in the emission line images. The `southern 
jet' is visible in the \halpha~and \sii~images, but it is found that almost 
all of the line 
emission originates from the `northern jet'.  Morphologically, the line 
emission images look fairly 
similar to each other. Major differences first become obvious when we 
analyse the working surface -- feature D. In Figure \ref{has2cont}
contour plots showing the 
detailed structure of the feature D in the light of these two ions, 
are displayed. The bright core seen in the R- and I-band 
images is obvious in the \halpha~image, but less so in the \sii~data. 
Instead in the latter band we see a mottled structure with a number of 
point-like emission centers with the brightest emission further 
{\em away} from IRS5 than the bright core seen in the \halpha~data (see also 
Figure \ref{diffcont}). Referring to the ground based data of FL94 it was also 
remarked upon that the \sii~intensity maxima were `further out' from 
IRS5. 

 Using images obtained under sub-arcsec seeing conditions, and 
then applying an unsharp mask, FL94 found a 
separation of \halpha~and \sii~of 0.5 arcsec. {\em This is confirmed by the 
present HST data set.}
\subsection{Spectroscopic results}
 In Tables \ref{ion1} and \ref{ion2} we present 
a summary of our spectroscopic results. The high spectral resolution data 
obtained towards the working surface (feature D), is presented in the 
first table. Here the peak radial velocity is found between 
-160\kms~and -185\kms. The maximum FWZI (Full Width at Zero Intensity) 
is \about~300\kms, which should be identified with the shock velocity 
at the apex \citep{hrh87}. This is also confirmed by the observation 
of \oiii~emission from the working surface by \citet{im05}.

 In Table \ref{ion2}, we show the medium resolution 
spectroscopy obtained through slits positioned such as to map 
out the radial velocity field of the two jets. Also in  Figure 
\ref{alfoscspectra}~we demonstrate both the positions where the 
spectraq were obtained, as well as samples of the spectroscopic 
information (\halpha~and \feii~7\,155\AA). These spectra clearly show 
our conclusions drawn about the velocity field in sections 3.4, 4.1 
and 4.2. From the spectroscopic data, it is clear, that there 
is indeed two velocity systems connected with the jet.  A comparison 
between the spectroscopy and the images allow a one-to-one 
correspondence between the identified features (F3 -- D), and the 
velocity components in the spectra. These results are presented in 
detail in Table \ref{ion2}. The highest 
radial velocities are found to be associated with the northern, 
brighter component, where essentially all of the visible shock induced 
knots are found. A lower (more positive) radial velocity system is 
connected with the southern component. The highest velocity found in 
any ion within the jet appear to be associated with the innermost 
features (F, F2), where peak velocities of 430 \kms, and FWZI of up to 
400 \kms are present (see also section 4.2). A comparison between our 
data and the IR \feii~spectroscopy of \citet{pyo02} is fully 
consistent when taking into account differences in sensitivity. 
A more thorough discussion 
about the spectroscopic results can be found 
in \citet{lfl05}~and \citet{im05}.
\subsection{The nature and proper motions of the clumps along the jets}
 With one major exception, all well defined features that can be 
identified in both HST epochs are to be found in the northern jet. The 
exception is the bright oval feature seen in the I-band 
(Figure \ref{wowjet}, see also discussion).

\begin{figure*}
  \begin{center}
    \leavevmode 
\includegraphics[angle=-90,totalheight=18cm]{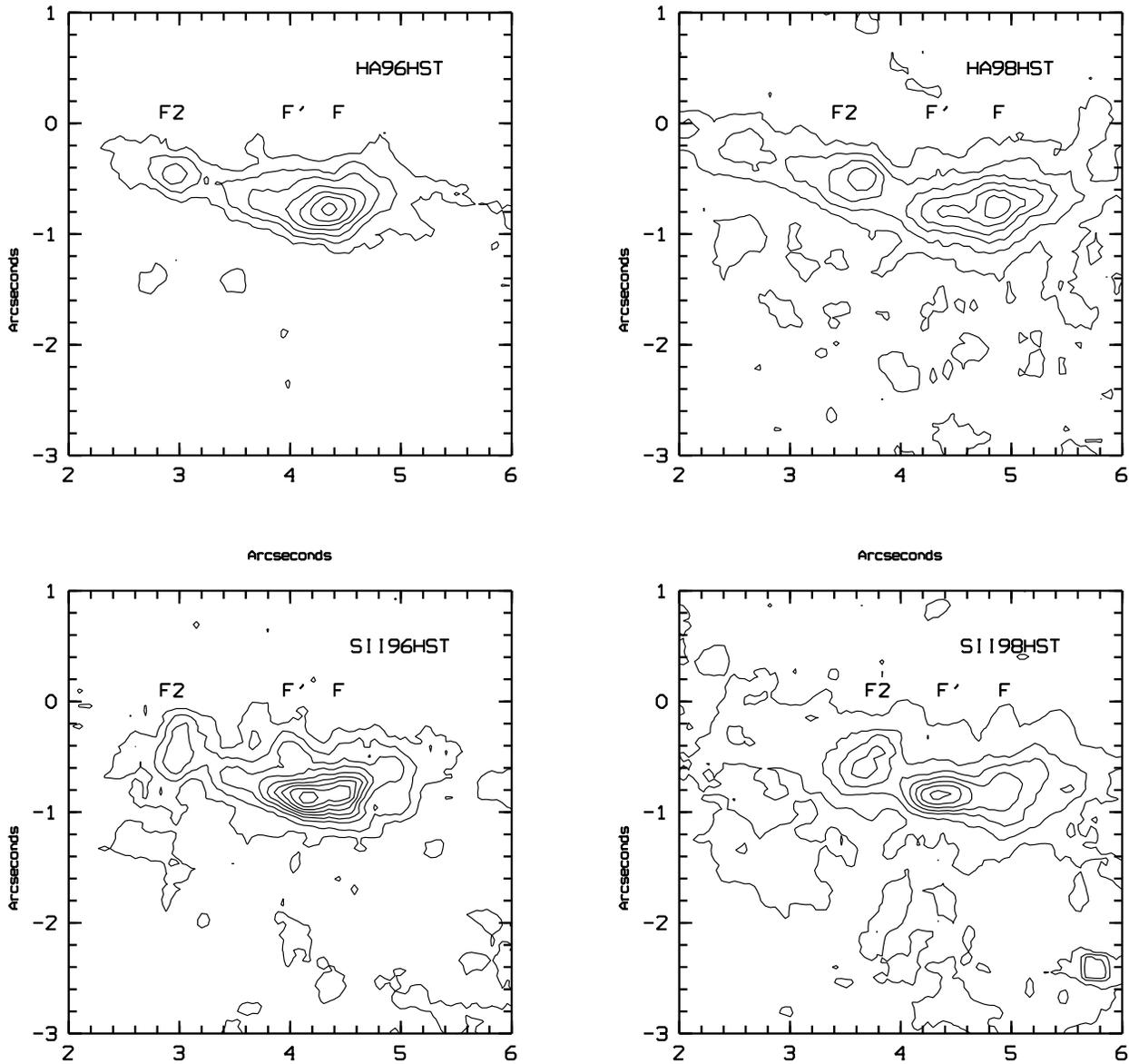}
    \caption{The 1996 (left) and 1998 (right) \halpha~(Top) and \sii~(Bottom) 
contour representations of the region around knot F.
North is up and East towards the left. The lowest contour is 
2$\times$~10$^{-18}$\ecs~pixel$^{-1}$, (3$\sigma$) and each consecutive 
contour is 
separated by 1$\sigma$. 
Note the proper motions over the 770 days between the 2 epochs 
representative of velocities of between 120 \kms~and 250 \kms~ if 
representative of actual motions.}
\label{fcomplex}
  \end{center}
\end{figure*}

 Proper motions of features in the 
jet were found by \citet{nec87}, and further studied by FL94. These 
ground based observations led to the identification of the features 
denoted A-F. Features A \&~B are found further out than the others 
and are thus not on the PC chip of the WFPC2. We refer 
to FL94 for illustration. In both \citet{nec87}~and FL94 it was found that distinct new knots appeared from 
behind the obscuration along the line of sight to IRS5. FL94 determined the 
transverse velocities for knots D, E and F to be between 120 
\kms~and \about~300 \kms. Knot E appeared 
well defined in the middle of the jet in the data from 1989, but in 1993 
it seemed on the verge of merging with the working surface D. At the 
same time there were indications of a new feature appearing at the 
original spot of E. When comparing the 
separate \sii~and \halpha~images, we find that the features 
appear to be well defined in each filterband image, but are found at 
somewhat different positions. This could be explained essentially 
through a change of excitation conditions at different positions (see 
Table \ref{pm1}~and Figure \ref{siicontour}~and 
\ref{fcomplex}). In some of the features there are indications of them 
being representative of clumps of gas moving at a high speed through 
the surrounding gas (so called {\em interstellar 
bullet}~\citep[e.g.][]{ns79}). This is 
especially true for knot E, which is compact and appears at different 
positions at the two HST epochs. This feature is, however, 
currently traveling at transverse velocities about half of those measured 
by FL94. This behavior could be explained by the 
present feature E being either:
\begin{itemize}
\item A different bullet -- ejected at a different velocity
\item The feature E observed by FL94 -- from the ground -- was 
the seeing-smoothed agglomeration of several "bullets"
\item A combination of both these options
\end{itemize}
%
%
\begin{table*}[htb]
\begin{center}
\caption{Proper motions measured for features in the jet. Note that 
these data are obtained only from the HST 1996 \& 1998 observations. The 
actual precision of measurement is 0.1 arcsec. Note that the 
velocity data for the F-complex is somewhat uncertain, since the 
identification of individual features (particularly F') have some ambiguity.}
\label{pm1}
\begin{flushleft}
\begin{tabular}{llllllr}
\hline
Feature & \halpha & \sii & Average & Velocity & Velocity@150pc & Comment \\
\hline
 & arcsec & arcsec & arcsec & arcsec yr$^{-1}$ & \kms & \\
\hline
F2 & \asecdot{0}{76} & \asecdot{0}{73} & \asecdot{0}{745} & 
\asecdot{0}{35} & 250 & See text about F-complex \\
F' & \asecdot{0}{6}& \asecdot{0}{7}& \asecdot{0}{65}& \asecdot{0}{3} 
& 220& difficult to identify unambiguously \\
F & \asecdot{0}{5} & \asecdot{0}{7} & \asecdot{0}{6} & 
\asecdot{0}{3} & 220 & See text about F-complex \\
E & \asecdot{0}{53} & \asecdot{0}{41} & \asecdot{0}{47} & 
\asecdot{0}{22} & 160 & \\
D2 & \asecdot{0}{65} & \asecdot{0}{62} & \asecdot{0}{635} & 
\asecdot{0}{30} & 215 & \\
D1 & \asecdot{0}{44} & \asecdot{0}{45} & \asecdot{0}{445} & 
\asecdot{0}{21} & 150 & \\
D4 & -- & \asecdot{0}{44} & \asecdot{0}{44} & \asecdot{0}{20} & 150 &\\
D3 & \asecdot{0}{36} & -- & \asecdot{0}{36} & \asecdot{0}{17} & 120 & \\
\hline
\end{tabular}
\end{flushleft}
\end{center}
\end{table*}
 As is evident from Table \ref{pm1}, the knots D2 and F2 (see 
Figure \ref{mypetjet})~are the knots with the highest velocities, consistent 
with the results of FL94. 

Our radial velocity results (see Figure \ref{alfoscspectra}) indicate 
that the highest velocities {\it for a given inclination}~should be found 
as far in towards IRS5 as possible, which is consistent with F2 having the 
highest transverse velocity (see Table \ref{pm1}). 
 In Figure \ref{fcomplex}, we have zoomed in on the area around knot 
F. A number of smaller, well defined (sub-)features are obvious in the   
images. This {\em F-complex} consist of sub-features F2, F' and 
F. The F2 knot is identifiable as a bright isolated knot, of 
intermediate excitation,  seen in both 
filters and in both epochs, while F' has low-excitation 
characteristics and is harder to identify unambiguously in all 
filters/epochs. Finally, the F feature is of higher excitation.

 The lengthening of the jet with time, as measured by its total length at 
each epoch is shown in Table \ref{length}, where we have collected 
all the data for this parameter that we have available. {\em The 
average change in length found during this period (1 Dec 1989 to 21 Oct 
2001 = 4\,342 days) is 0.3 arcsec $yr^{-1}$~representative of a transversal 
velocity of 235 \kms}.
%
%
%
\begin{table}[htb]
\begin{center}
\caption{The length of the jet at different epochs. $\Delta$~is the 
number of days between successive epochs. The measurement precision is 
\about~0.1 arcsec.}
\label{length}
\begin{flushleft}
\begin{tabular}{lllll}
\hline
Filter & Year & $\Delta$t& Length & Rate of change\\
& & Days & Arcsec & Arcsec $yr^{-1}$ \\
\hline
R & 1989 & 0 & \asecdot{10}{4} & na \\
R & 1993 & 1364 & \asecdot{11}{0} & \asecdot{0}{16} \\
R & 1996 & 935 & \asecdot{12}{24} & \asecdot{0}{48} \\
\halpha~+ \sii& 1998 & 727 & \asecdot{12}{73} & \asecdot{0}{25} \\
\halpha & 2001 & 1316 & \asecdot{14}{32} & \asecdot{0}{44} \\
\hline
\end{tabular}
\end{flushleft}
\end{center}
\end{table}

 In Table \ref{abc}, we compare the velocity in the plane of the sky 
of the knots A, 
B, and C (see FL94 for location) over total timespan covered by the 
data with the results in our earlier paper. There is maybe an indication that the 
knots are slowing down as they progress further out beyond the working 
surface, as well as becoming somewhat less well defined. This needs to 
be confirmed with observations over a longer timebase. Nevertheless, 
these features have now existed as well defined entities since at 
least 1983 \citep{nec87}. In the meantime they have moved between 600 
and 1\,000 AU further downstream. 
%
%
%
\begin{table}[htb]
\begin{center}
\caption{The change of distance of features A, B and C}
\label{abc}
\begin{flushleft}
\begin{tabular}{lll}
\hline
Feature & \citet{fl94} & 1993 -- 2001 \\
& Arcsec $yr^{-1}$ & Arcsec $yr^{-1}$ \\
\hline
A & \asecdot{0}{29}$\pm$\asecdot{0}{06} & 
\asecdot{0}{1}$\pm$\asecdot{0}{1} \\
B & \asecdot{0}{26}$\pm$\asecdot{0}{06} & 
\asecdot{0}{18}$\pm$\asecdot{0}{1} \\
C & \asecdot{0}{26}$\pm$\asecdot{0}{05} &
\asecdot{0}{20}$\pm$\asecdot{0}{1} \\
\hline
\end{tabular}
\end{flushleft}
\end{center}
\end{table}
 The northern jet appears to be wiggly -- 
particularly in the \halpha~band image. The amplitude of the 
excursions is 
\about~0.5 arcsec, and the distance between the nodes is 
3 arcsec~or \about~450AU (\about~7$\times10^{15}$ cm). In unpublished 
ground based data from 1996 (\halpha), we have used an unsharp mask on 
the data, and thanks to a much higher signal-to-noise ratio confirmed 
the size and extent of the pattern.

 FL98 interpreted the HST data from 1996 as indicative of at least 
two physically separated velocity components of the jet. In our new 
spectroscopic observations we have complete coverage of 
the jet with medium and high resolution spectroscopy, which 
allows us to determine the velocity field of the entire jet 
(Figure \ref{alfoscspectra}). From 
different line ratios we can determine some of the physical 
parameters.  In order to do so, we use previously unpublished data, 
where we have imaged the whole SW L1551 complex through narrow interference 
filters. Those filters included \halpha, \hbeta~and 
the \sii~4067/68\AA~ion. We can then a) deredden our data and b) determine the 
electron temperature and electron density along the jet, and in and around 
the working surface (knot D).
 The electron density, \elec, can be determined 
from the \sii6717\AA/\sii6731\AA~ratio, and a suitable model. We use the 
same recent 5-level model for \sii~as \citet{flg98}. The electron 
temperature, $T_e$, can be determined from the same data together with 
information about the flux from the \sii~4067/68\AA. Although 
inherently faint, both knot F and knot D are detected in the blue 
\sii~lines, and with fluxes of 1.25 
$\times$~10$^{-16}$~\ecs~(S/N \about~4) and 7.05 
$\times$~10$^{-16}$~\ecs~(S/N 
\about~11). Note that the apertures used are different. 
Dereddening is achieved from the \halpha/\hbeta~ratio (see 
Tables \ref{mytable}~and \ref{exttable}). Using the above 
mentioned spectroscopic data, 
and our \sii~model, we arrive at values of $T_e$(D) \about~8\,700 
K$\pm$1\,000 K, $T_e$(F) \about~14\,250 K$\pm$1\,750 K, $n_e$(D) 
\about~1\,600 \cmthree ~$\pm$~200 \cmthree, 
and $n_e$(F) \about~1\,800 \cmthree~$\pm$~200 
\cmthree. Note that these values then are averages 
over the aperture used in each case. For knot F this is about 300 
$\times$~300 {\em AU}$^2$, while for feature D it is \about 530 
$\times$~530 {\em AU}$^2$, assuming a distance of \about~150pc.

 We have also analysed our \feii~data using a model atom including 142 
levels with 1\,438 transitions. Our model incorporates radiative 
transfer, but in this case this is unnecessary since the optical 
depth in the \feii~lines is of order $10^{-8}$. We have detections of 
 \feii~7\,155.1\AA, 7\,172.0\AA, 7\,452.5\AA~and 8\,619.3\AA~for knot 
D, and our best fit indicates higher $T_e$~and $n_e$ than the \sii~data, viz. 
13\,500K and 3\,200 \cmthree, respectively. Given the somewhat different 
apertures through which we are averaging our data, the values are quite 
reasonable, however. {\em A puzzling result, however, is the absence 
of the \feii~line at 6\,875.8\AA.} In our model, which assumes a 
solar abundance, the line at 7454.6\AA~should always be weaker. We 
note, however, that in observations of HH1, the same situation occurs, 
i.e. the detection of 7454.6\AA,  8\,619.3\AA, but 
not 6\,875.8\AA~\citep{solf88}. 
This suggests an explanation. Like in the case of HH1, we do not detect 
any permitted FeII lines. As pointed out by \citet{solf88}, the 
permitted FeII lines require an excitation energy of, on average, 
5.2 eV, while the forbidden lines have lower excitation energy. All of 
the detected lines in our spectrum have excitation energies of 1.7 -- 
2.0 eV, while the 6875.8\AA~line, falls in between since it 
requires \about~3.8~eV. It appears 
thus that the kinetic energies in knot D are not high enough to 
excite forbidden lines with higher energies than \about~2.0 eV. There 
is no indication of any line with a higher excitation energy than 
that, although a marginal detection of 6875.8\AA~is found at the 
location of knot F, and at a radial velocity of \about~-300 \kms. An 
alternative explanation is that the atomic data, as given 
in \citet{q96}, for the 6875.8\AA~line are in error.
\subsection{ The extinction towards the jet}
In Table~\ref{mytable} we present the flux from some of the persistent 
features found along the jet as a function of time (including some of 
our unpublished data). The presence of \hbeta~emission allows 
us to determine the extinction towards separate parts of the jet 
(assuming the appropriate recombination case), where one would 
expect the extinction to grow as one progresses towards the position 
of IRS5. 

 \citet{im05}~have detected \oiii~emission that can be taken as direct 
evidence of a strong shock being present at the position of knot 
D. Tentatively, we then identify the feature D 
as a (relatively) strong shock. Somewhat weaker \oiii~emission is 
also detected from the position of knot F. It is at these positions we then 
expect that the 
collisional contribution to the population of \halpha~to be 
insignificant \citep{hrh87}. We can thus use the Balmer decrement 
for case B recombination in order to estimate the extinction. 

 As can be seen from Tables \ref{mytable}~and~\ref{exttable}, in 
1996 we find a ratio of 
\halpha/\hbeta~of 8.5 for feature D, while the data 
obtained in 2001 indicate a ratio of 7.0 for the same object. During the 
intervening time, this  feature has moved away from IRS5 at a rate of 
\about~0.2 arcsec~yr$^{-1}$~(Table \ref{pm1}).
\citet{cf85}~found a \halpha~to \hbeta~ratio of 
\about~25. This value, however, should be used with some caution, 
since we note that they observed using blind offsets from a nearby star, and 
the \halpha~and the \hbeta~spectra were obtained at different times 
(pointings). There could thus be an offset by an unknown 
amount between \halpha~and \hbeta. Further, our data indicate that the 
proper motion of this feature over \about~13 years, would implicate a 
displacement of \about~2 -- 3 arcsec, which means that the physical 
conditions could be very different.

 Using a normal galactic extinction curve 
\citep{rl85}~our data for knot D indicate A$_V$~\about~3 magnitudes 
in 1996 and \about 2 magnitudes in 2001. A $\theta$~Orionis extinction law 
\citep{cc88}~gives values of 3.6 and 2.9 magnitudes of 
visual extinction respectively. Applying the data of Cohen and 
Fuller (1985 -- data obtained 1982) we have a visual extinction of 5.8  and 
7.4 magnitudes depending on the extinction law applied.

 As we progress further in along the jet, towards IRS5, the 
extinction increases. It is close to 4 magnitudes at the position of 
feature F. Towards the very tip of the jets, i.e. the apex where they 
suddenly emerge from invisibility, we find that the ratio is indicative of 
an extinction of 4 to 6 magnitudes. These values are consistent with the 
extinction calculated from the X-ray spectrum for the L1551 IRS5 
source \citep{fav02}.

\section{Discussion}
\subsection{The working surface and the physical conditions along the jets}
 The first, most easily accessed, physical parameter determined from our HST 
data is the level of excitation as defined by the ratio of 
\sii~flux to \halpha~flux \citep{rag96}. For the purpose of this paper 
we will call this ratio "the excitation index". To display 
the positions where the low excitation \sii~emission dominates over 
the \halpha~emission -- and vice versa we have computed the difference 
images between \halpha~flux and \sii~flux (e.g. Figure \ref{diffcont}). 
In this figure the most obvious feature is the end of the jet (Feature 
D) where 
prominent \sii, representative of low-excitation emission and displayed in 
white is wrapping around the prominent \halpha~(displayed in black) 
which is representative of high-excitation emission. Since we know the 
direction and rough inclination of the jet, the morphology leads to the 
interpretation as being a nice example of a working 
surface with the bow-shock (strong \sii) wrapping itself around the Mach-disk 
(strong \halpha). 

 Taken together with the excitation index, as defined above, the ratio 
of \oiii/~\hbeta~refines the understanding of the excitation conditions. 
According to \citet{rag96}, a value of \oiii/~\hbeta~larger than 
0.1 together with a \sii/\halpha~smaller than 1.5 indicates high excitation, 
while opposite values are indicative of very low excitation conditions. 
\citet{im05} have found \oiii~5\,007~\AA~emission emanating from both 
features F and D, as well as \hbeta~emission distributed in the same way as 
the \halpha~emission (see discussion about the extinction above). 

 Feature F is found by \citet{im05} to have an 
\oiii/\hbeta~ratio of 0.3$\pm$0.06, while the ratio for 
knot D1 is 0.22$\pm$0.02, both ratios measured in 2001. These are 
values indicative of a high level of excitation. For the 
\sii/\halpha~ratio we have data covering a larger time span. 
Feature F (see Figure~\ref{mypetjet}) has an \sii/\halpha~ratio 
varying from 0.6$\pm$0.1 in 1994 to 1$\pm$0.1 in 1998, thus indicating that 
an initial high excitation feature is somewhat weakening with time. An 
analysis of the F-complex 
in the HST data shows that the 3 features F2, F and F' had 
\sii/\halpha~ratios of 1.0$\pm$0.1, 1.9$\pm0.2$ and 0.9$\pm$0.1 in 1996. In 
1998 the values for the ratios were 1$\pm$0.1, 0.9$\pm$0.1 and 
0.8$\pm$0.1 respectively. The feature F as observed 
from the ground, is of course the agglomeration of all three 
subcomponents. {\em These observations demonstrate the occurrence of changes 
in the jet not only with respect to the morphological properties, but also 
to the physical conditions within a time scale of a few years}.
Knot F is also the second brightest emission feature 
along the two jets. Careful analysis of the HST data show that 
a very high level of excitation is mainly found along the outer 
(northern) edge of feature F, while a very thin 
(\about~a few pixel thick) layer of lower excitation (demonstrated by 
strong \sii~emission in our difference images) is found along the 
southern edge of the jet. Taken together with 
the \oiii/~\hbeta~index mentioned above we conclude that (at least 
parts of) knot F is of high but variable excitation in the classification 
scheme of \citet{rag96}.

 The fainter feature E has an \sii/\halpha~ratio of 0.6 both in 1996 and in 
the 1998 data. This is thus indicative of a relatively high level of 
excitation, {\em regardless of the fact that the feature has 
moved a significant distance in the intervening time.} It is, however, 
not detected in \oiii~by \citet{im05}.

\begin{figure*}
  \begin{center}
    \leavevmode 
     \includegraphics[angle=-90,totalheight=18.0cm]{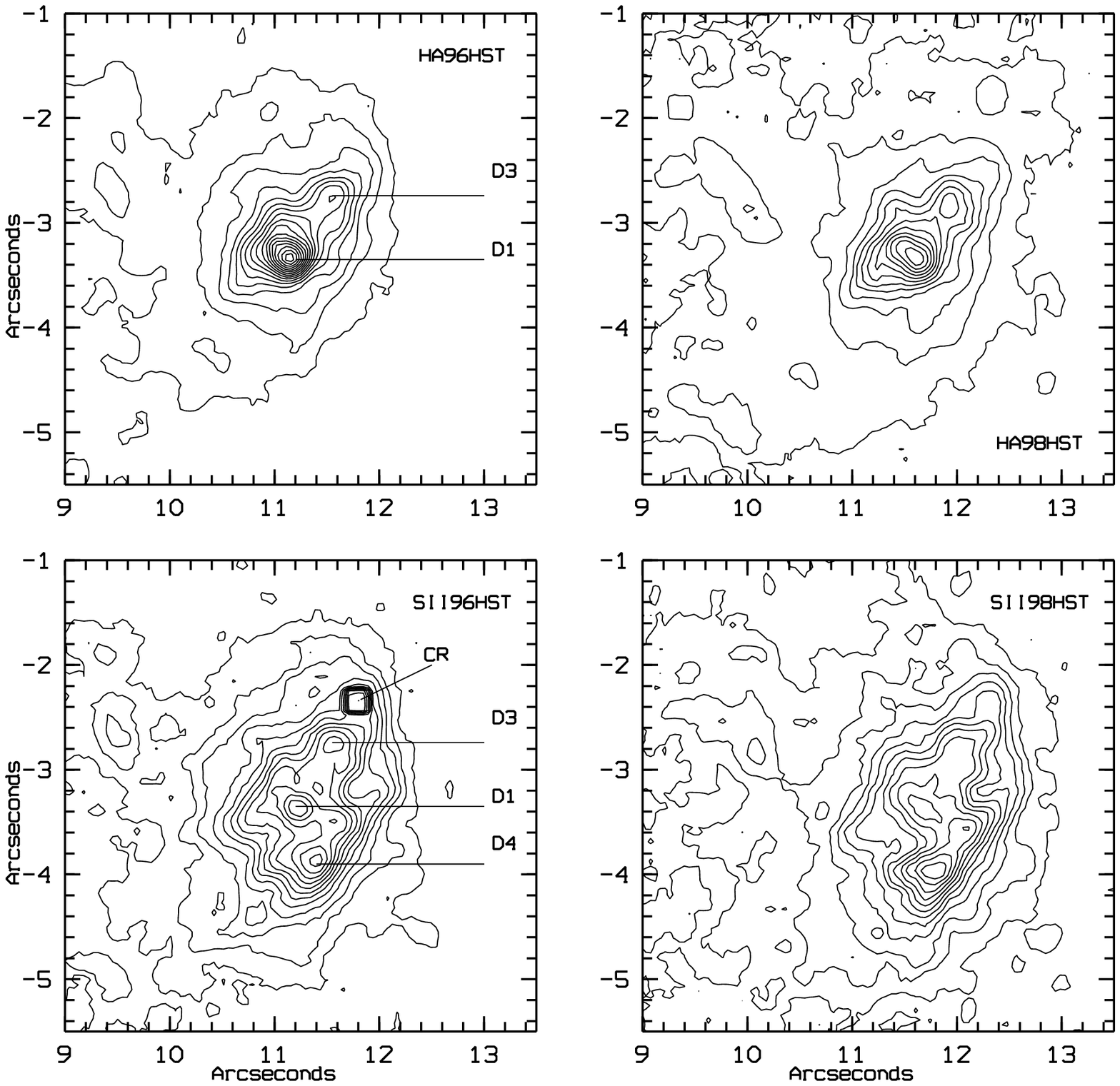}
    \caption{The 1996 and 1998 \halpha~(top) and \sii~(bottom) contour 
representations of feature D -- the presumed working surface. 
North is up and East towards the left. The lowest contour is 
2.5$\times$~10$^{-18}$\ecs~pixel$^{-1}$, and each contour is 
separated by 1.0$\times$~10$^{-18}$\ecs~pixel$^{-1}$. The bright, compact 
feature, dominating 
the \halpha~frame has been given the designation D1 in this paper. 
Note the proper motions over the 770 days between the 2 epochs 
representative of transverse velocities of about 150 \kms. CR is a 
cosmic ray strike.}
\label{has2cont}
  \end{center}
\end{figure*}

\begin{figure*}
 \begin{center}
    \leavevmode 
    \epsfig{file=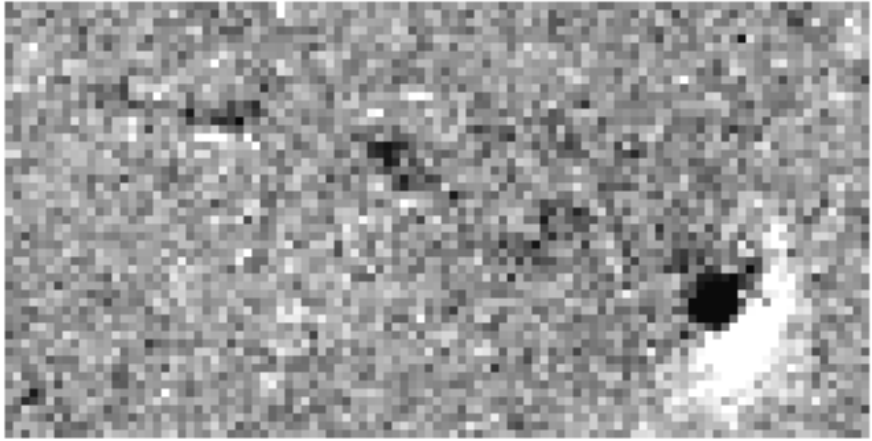, width=14.5cm, bbllx=120pt,
      bblly=380pt, bburx=570pt, bbury=670pt, clip=}
    \caption{An \halpha~-\sii6717+6731\AA~difference image (1996 data) 
of the complete jet, shown as a greyscale. Here \halpha~is represented by 
black, while the \sii~is white. The picture is smoothed to 0.1 arcsec 
resolution, and the image is 11 arcsec by 6 arcsec. 
The strong white/black feature is knot D -- the working 
surface. The spatial separation of black (\halpha) and white(\sii) is 
interpreted in the text as the resolving of the shock into the Mach 
disk (black) and forward- or bow-shock (white). Note that apart from the 
working surface the rest of the jet cancels out almost perfectly, 
leaving only a few filaments of separated emission.}
\label{diffcont}
  \end{center}
\end{figure*}

Finally, D, the working surface, 
is found to have an average level of excitation index of \about~1, with very 
little variation over time (see Table \ref{mytable}). If we analyse the 
separate, very bright knots that make up what we have been calling 
feature D, we find a more complex picture. In the \halpha ~and 
the \sii~images (see Figure \ref{has2cont}), we see a number of bright 
(sub-)features. The brightest of these are the ones where we argued in FL98 
that they were the signature of the working surface of the jet with the Mach 
disk brightest in \halpha~and the forward shock 
(or bow-shock) being brightest in \sii~respectively. They are now designated D1 (brightest in 
\halpha) and D4 (brightest in SII). D1 has an excitation index of 
\about~0.5 making it a high excitation object. This is also supported 
by the \oiii/~\hbeta~ratio of 0.22 (see above). The opposite is true 
for D4 as well as for the `bow-shape' surrounding the core. Here we find 
excitation indeces of \about~1 -- 5 indicating low-excitation conditions. 

\begin{figure*}
 \begin{center}
    \leavevmode 
	\includegraphics[angle=-90,totalheight=5cm]{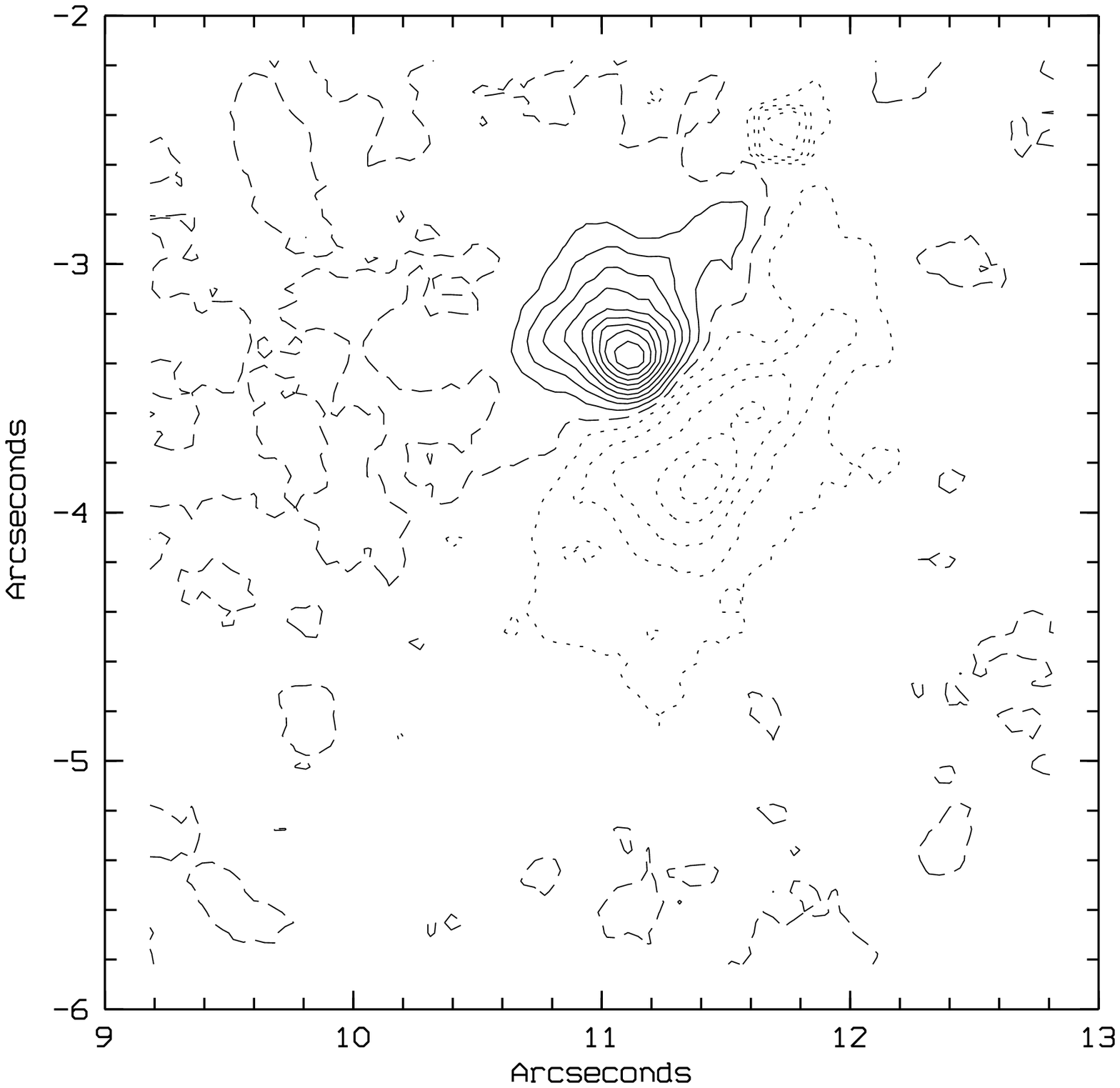}
	\includegraphics[angle=-90,totalheight=5cm]{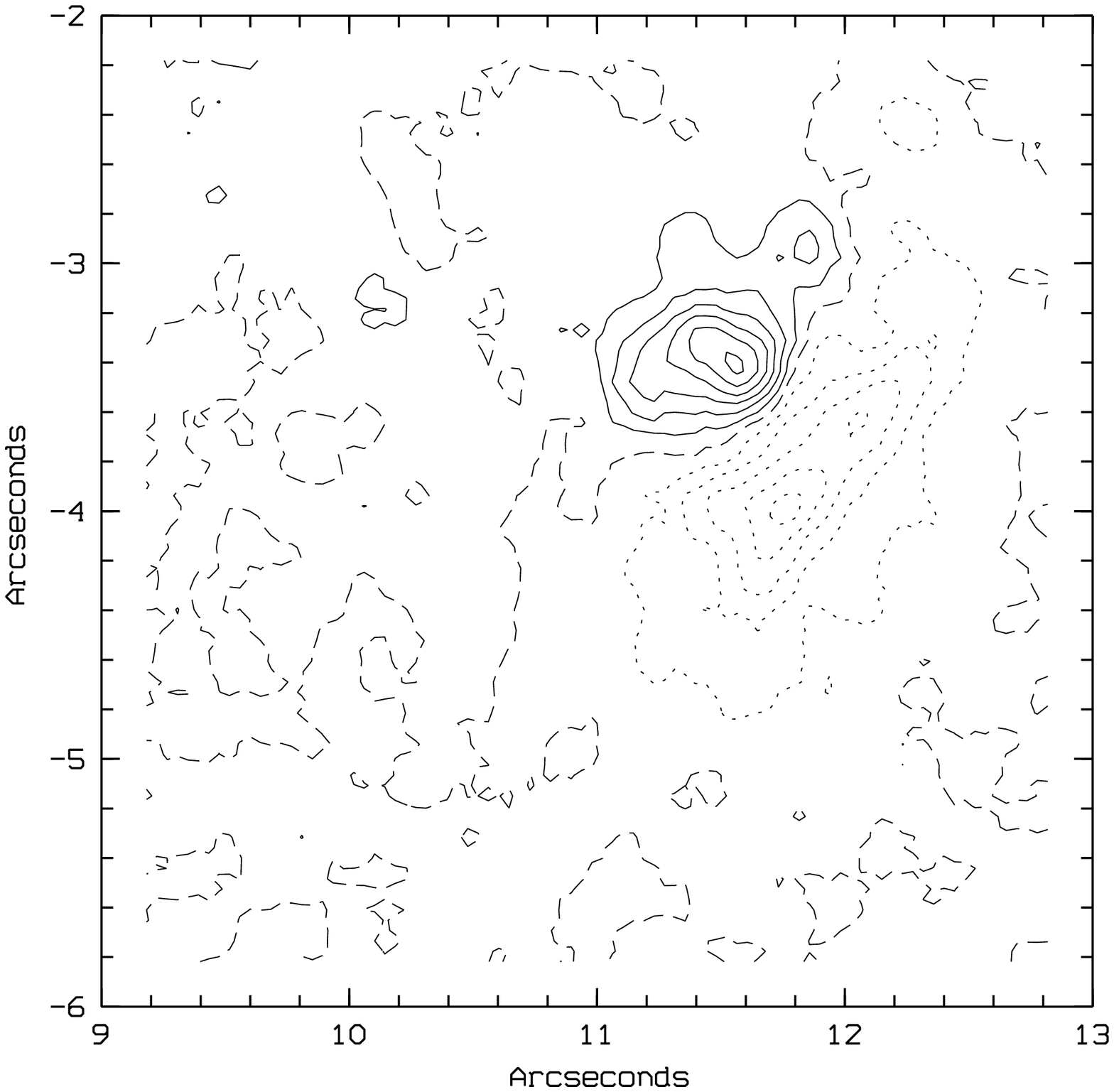}
    \caption{Difference images of the working surface (feature D -- see text) 
between the \halpha~ and \sii~images in 1996 (top) and 1998 (bottom). 
The positions are relative to an arbitary position.}
\label{diffjetjet}
  \end{center}
\end{figure*}

\begin{figure*}
   \centering
	\includegraphics[totalheight=3cm]{2399fi11.eps}
	\includegraphics[totalheight=3cm]{2399fi12.eps}
	\includegraphics[totalheight=3cm]{2399fi13.eps}
	\includegraphics[totalheight=3cm]{2399fi14.eps}
	\includegraphics[angle=-90,totalheight=14.5cm]{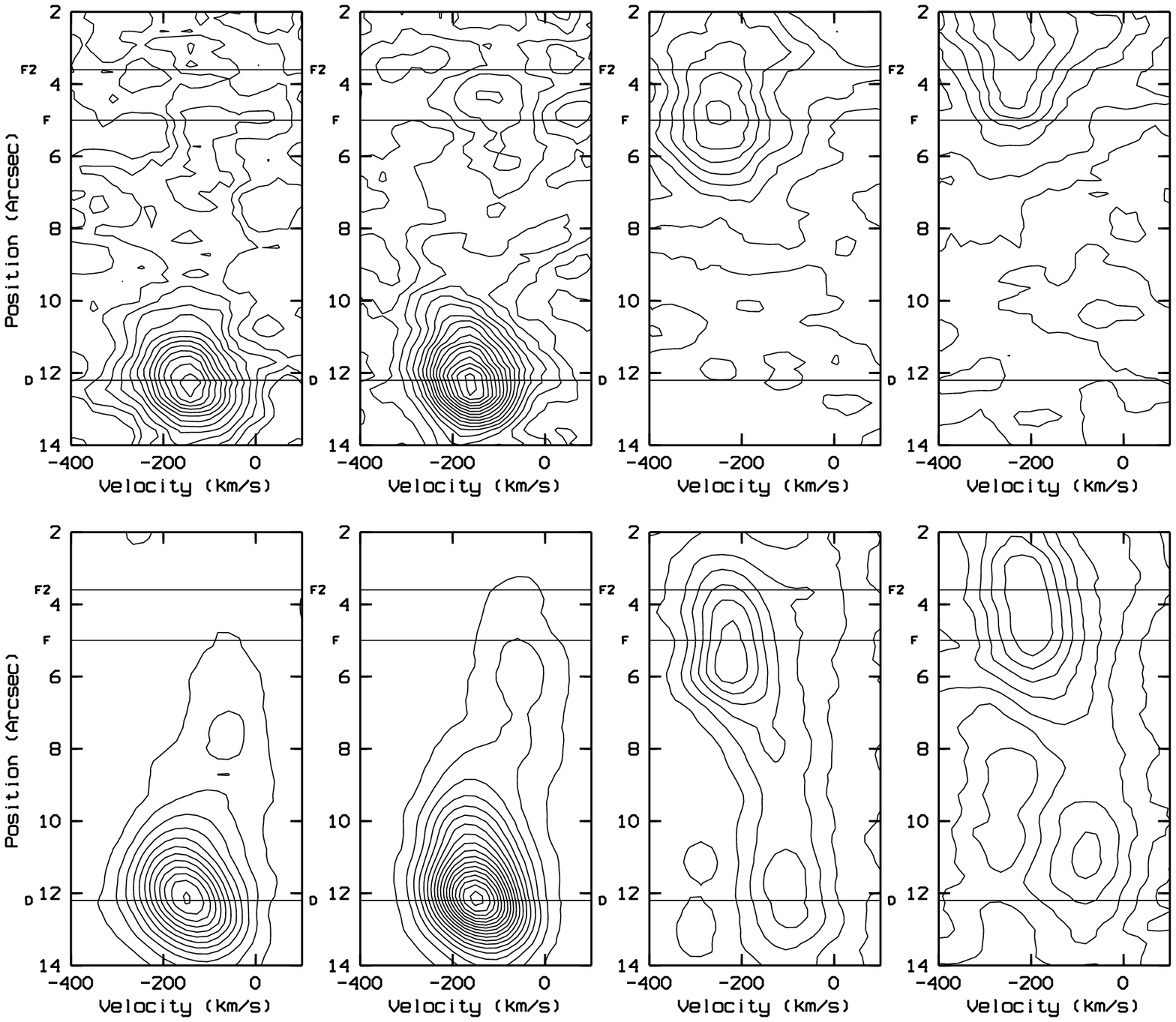}
	\caption{The \halpha~and \feii~7155\AA~spectra obtained with 
ALFOSC. In the top row, the different slit positions are shown superposed 
on 20s R-band images. The effective slit width is 12\asec~long.
In the middle row are the \feii~spectra and in the bottom 
rows are the \halpha~spectra. All are displayed as contour maps, where 
the contours have been selected to show the maximum extent of 
emission. A radial velocity scale w.r.t. barycentric zero has been 
used. Comparing on the one hand the spectra of \halpha~and 
\feii~7155\AA~spectra and on the other the slit positions with repsect 
to the jet, clearly demonstrate the two different velocity systems, 
originally discovered by FL98. At position 12.2\asec, in both spectral 
lines, feature D is prominent. A peak velocity of 160 \kms is indicated. 
Features D, F and F2 have been marked according to their 1998 HST 
positions}
\label{alfoscspectra}
\end{figure*}

 We already identified the entire feature D as the working surface 
of the jet in FL98. The characteristic features allowing us to make this 
identification are clearly visible in Figure \ref{diffcont}. Also 
the interface between the regions dominated by \halpha~and \sii~emission is 
clearly discernible in Figure \ref{diffjetjet} which shows contour 
plots of the two HST epochs and where the \sii~emission has been subtracted 
from the \halpha~emission. The latter emission is strongly 
concentrated towards 
the inner part of the working surface, at the position of the high 
excitation core, while 
the \sii~is `swept' into a nice bow-shape surrounding it. The steepest 
gradient of the \halpha~emission is on the side {\em away}~from IRS5 
and on the interface with the region dominated by \sii. Comparing with 
the individual images of \halpha~and \sii~in Figure 
\ref{has2cont}~we see that the \sii~have its steepest gradient 
where the bow-shape is furthest {\em away} from IRS5. {\em In the context 
of the model of \citet{har89}, we interpret these gradients to represent the 
locations of the reverse shock or Mach disk (\halpha), and the forward, or 
bow-shock (SII).} The bright \halpha~core of the working surface has a 
FWHM of 0.3 arcsec~at right angles to the major 
axis of the outflow, while the whole `working surface', knot D, has a 
FWHM of \about~2 arcsec. The separation between maximum \sii~(D4) and 
maximum \halpha\ (D1) emission is 0.6 arcsec~or about 2~$\times~10^{15}$ 
cm deprojected for a viewing angle of 45\adeg. We are thus resolving the 
structure of the shock. 

 {\em This separation is 
consistent with the calculations of \citet{har89} that -- given a 
cooling distance of \about~$R_{jet}$~(= the jet radius) --  show that 
both the forward shock, i.e. the bow-shock, and the Mach disk will be radiative 
and that the separation should be just equal to \about~$R_{jet}$, as 
is true in this case.} In our data both the cooling length -- i.e. 
the distance between D1 and D4 -- and the 
jet radius appear to be of the same order (\about~10$^{15}$~cm). According 
to the model of \citet{har89}, a Mach disk 
which is -- as in this case -- significantly brighter and here also with a 
much higher level of excitation than 
the bow-shock \citep[see][ also Fig.~3]{fhl97} is indicative of a 
less dense jet penetrating into a denser ambient medium. It is at this 
interface where it is possible that the transfer of momentum from the fast 
atomic wind to the slower molecular outflow is taking place. 

 The model of \citet{har89}~predicts the relative density of the jet 
and the ambient medium surrounding the jet in terms of the relative 
brightness of the bow-shock and the Mach disk, and the jet velocity. 
For the HH154 jet, we find the density of the jet:

\begin{equation}
 \rho_{jet} \sim (0.05 - 0.1) \times \rho_{ambient} 
\end{equation}

 An analysis taking into account the \elec~derived in the previous 
section and the molecular data of Fridlund \& Knee (1993) 
then shows that the ionisation fraction of the jet is \about~1 
(FL98). This led to the conclusion in that paper that the optical jet 
is not driving the molecular outflow, since there is a lack of the 
necessary momentum by at least ~2 orders of magnitude. In FL98, however, we 
assumed a velocity of 300 \kms~for the jet material (since we then 
did only have access to low dispersion spectroscopy). Inspection of 
Figure \ref{alfoscspectra}, demonstrates that most of the material is 
found at velocities {\em less}~than this value. Some 
material is traveling at velocities close to 600 \kms, when taking the 
inclination into account, but we note that even if we were to 
allow the bulk of the material to travel at this higher velocity, it 
would still leave us lacking at least one order of magnitude of 
momentum (something which would also require a 100\% efficient mechanism 
for the transfer of the jet momentum to the molecular outflow). In this 
context, we also note that there are no obvious indications of so called 
`turbulent 
entrainment', and that further the interface for `prompt entrainment' (i.e. 
at the Mach disk) is very small -- only \about~30AU across. In this context, 
we wish to note, that the molecular momentum calculated by 
Fridlund \& Knee (1993) refers to the present outflow, i.e. with a dynamical 
time scale of less than \about~a few hundred years  which is therefore 
directly comparable to the jet dynamical time scale.
 
 In the \halpha-\sii~difference images in Figure \ref{diffjetjet}~we 
see indications of a change of position of the Mach disk. If we assume 
that the location of this shock is the sharp interface between the 
\sii~and the \halpha~emission, we can see that it has moved 
(0.5 arcsec) further out from IRS5 in \about~ 2 years. This corresponds to 
a velocity of \about~180 \kms, assuming a distance of 150pc. There is also an 
indication that the actual apex of the shock is turning towards the 
north -- something that could indicate rotation of the jet material.

\subsection{The location of the X-ray source}
Favata et al (2002) discovered 1 -- 2 keV X-rays emanating from the jet. 
Using {\em XMM-Newton}, the positional uncertainty included most of the 
jet. Nevertheless, these authors suggested that the 
X-rays were possibly emanating from the working surface in view of it being 
identified as a strong shock (FL98). This was in spite of it 
having a lower radial velocity and proper motion (both \about~160 \kms) 
than features found closer to IRS5 along the jet. \citet{fei03}, using the 
Chandra X-ray observatory, have improved on the position, and indeed locate 
the X-ray source in the innermost part of the jet (\about 1 arcsec~from 
the radio position of IRS5). As can be seen from Figure 
\ref{alfoscspectra}~in this paper, at this position we find radial velocities 
for features F \& F2 of up to 430 \kms, which given the 
assumed inclination of the jet 
of \about~45\adeg -- \about~60\adeg~\citep{lf05}~represent true velocities 
of \about~500 \kms~to \about~600~\kms. The immediate post-shock temperature 
is \citep{raga89}:

\begin{equation}
T_{ps} \simeq \frac{2.9 \times 10^5K}{1 + X} \times 
\left( \frac{v_{shock}}{{100~km~s^{-1}}} \right)^2
\end{equation}

Here X is the hydrogen preionisation fraction and an abundance of 0.9 (H) and 0.1 (He) by number has been assumed.
As mentioned before, FL98 have come to the conclusion, that it is very 
likely that the ionisation fraction
is close to unity.

With these assumptions, we calculate a shock temperature 
of \about~3 $\times$ 10$^6$~K to \about~10$^7$~K, in excellent agreement 
with the X-ray results of \cite{fav02}~and \cite{fei03}. We also note 
here that the material detected at the highest radial velocities 
probably emanates from the recombination/cooling region 'downstream' 
of the X-ray generating shock, where in all likelihood the velocity 
has dropped somewhat. All this 
information taken together therefore implies that the 
emission found at the highest radial velocities in our data, emanating from 
inside of the feature F2, is associated with the same location (or very close 
to) the source of the 
X-rays. Thus where the jet first becomes visible as it emerges from 
behind the obscuration. As can be seen from Table \ref{exttable}, the extinction toward 
feature F2 is compatible with the values determined from the XMM 
spectrum \citep{fav02}.

 We note here that the presence of an X-ray source at a distance of 
\about~200\,AU~from the surface of the disk, as well as other shock 
phenomena generating Lyman photons up to maybe 1\,000\,AU~to 
1\,500\,AU further away, {\em are going to have an impact on the 
ionisation structure of the disk surface}. This, in turn, could have an 
impact on models trying to explain the molecular outflows, such as MHD winds 
lifting off the disk surface.
\subsection{The bright feature in the inner region of the jet}
Both the two jets and the northern arc first appear in the same 
location in the sky ('apex') and have an opening angle of 
about 90\adeg (see Figure \ref{wowjet} -- the angle between the northern arc 
and the southern jet at their intersection). The jet emission is seen 
superposed onto a fan-shaped nebulosity that is most evident in the 
I-band filter. The northern arc and the southern jet then 
bend inwards towards the northern jet -- the northern arc more 
drastically than the southern jet -- and after 1.4 arcsec all 3 
components are parallel. {\em Note that it is within this envelope 
that X-ray emission is found to 
originate \citep{fei03}.} 

Absolute positional information in the HST data is
restricted to an accuracy of \about 1 arcsec, but our images are consistent
with the radio position of IRS5 \citep{loo97}~being within 0.5 arcsec from 
the origin of the visible jet. Only 0.6 arcsec SW 
from the apex is a very bright feature (see Figure \ref{wowjet}~bottom) visible 
in the I-band. It is also faintly visible in the R-band image, but is 
actually the brightest feature in the I-band. This 
knot appears elliptical with dimensions of about 
0.4 arcsec~by 0.2 arcsec~(FWHM). A careful analysis shows it to   
consist of two components, with a separation of 0.145 arcsec~or 
\about~20AU at the distance of L1551. An approximate $R-I$ colour for 
both sources is found to be \about~+2.0 magnitudes. 
 The feature is located between 
the northern jet and the southern jet, and at an 
angle offset from the outflow direction of those two 
features. \citet{white2000}, have analysed near-infrared HST-NICMOS data 
covering continuum and emission lines between 1.12$\mu$m and 
2.12$\mu$m. 
Inspection of these frames confirm the interpretation here, since 
faint \feii~emission is found from knot F, while a strong continuum 
peak is visible at the same position as the oval feature described 
here. No \molha~emission is detected in the NICMOS data.

 Since the jet emission is mostly caused by recombination -- 
presumably through interactions of the outflowing gas through shock 
phenomena, and since the oval feature described here is totally absent in 
the \halpha~and \sii~images we 
suggest that it is instead caused by light from the 
IRS5 system scattered up through a polar hole in the circumstellar 
disk and reflected off the inside of the `jet-cavity'. Its dimensions 
of \about~60AU by 30AU, with two peaks separated by \about~20AU, is 
also consistent with the binary nature and parameters of the 
IRS5 system. This is supported by the fact that \citet{rod03a,rod03b}
have demonstrated that the jets can be traced in towards the binary 
behind the obscuration, and indeed it thus appears likely that they 
together excavate a 'polar hole' through which the light from the 
stars/accretion disks can escape.

\section {Conclusions}
In summary, we find:
\begin{itemize}

\item We suggest that the bright, oval feature detected close to the apex of 
the jet in the I-band is caused by light from the 
IRS5 binary system scattered up through a (polar?) hole in the circumstellar 
disk and reflected off the inside of the `jet-cavity'towards our direction. This is supported by the colour of the feature.

\item We find that the position of the origin of the X-rays emanating 
from the jet is coincident or very nearly so with the location of the 
observed maximum radial velocities velocities (representative of true 
velocities of $\sim$~500~\kms -- 600~\kms).

\item We have mapped out the proper motions resulting in transverse 
velocities, as well as mapping the complete radial velocity field of 
the jet. 
A few jet features 
(e.g. knot E) now have very much different velocities. We suggest this 
is because the old feature has disappeared and we are actually 
observing a new "interstellar bullet" traveling down the jet towards 
the working surface. The time scale for the ejection of bullet like 
features is thus of the order of a few years.

\item The average rate of change of the length of the whole jet, found during 
our observing period (1 Dec 1989 to 21 Oct 2001 = 4342 days) 
is 0.3 arcsec $yr^{-1}$~representative of a transverse velocity of 235 \kms.

\item The time scale for variations of physical conditions within or 
near the jet (like the excitation conditions or the extinction) is of 
the order of a few years. This is similar to the time scale for 
morphological changes.

\item Within the  F-complex of knots, we find a number of well defined 
moving features (F, F' and F2), with large proper motions, but with 
much different excitation conditions among them.

\end{itemize}
 Further, we also find:
\begin{itemize}

\item The excitation conditions and the morphology within the feature 
D identify it as a "working surface" where the jet impacts either the ambient 
medium or material moving at a slower speed (from a previous ejection 
event). 

\item In the latter context, it is noteworthy, that feature D has 
maintained its individual character during more than 20 years and 
while traveling more than 1\,000 AU further downstream.

\item The Mach disk and the forward (or bow-)shocks are well resolved. 
The separation between high excitation (as traced by \halpha), 
and low-excitation (as traced by \sii~in the working surface (knot D) 
is \about~0.6 arcsec.

\item The Mach disk is moving 'downstream' with a 
transverse velocity of $\sim$~180 \kms. During the time interval 
between our HST observations it also appears to have changed 
direction. 

\item  The cooling distance in the working surface of the northern jet 
is \about~equal to the jet radius, $R_{jet}$~\about~0.6 arcsec

\end{itemize}
%
%

\begin{acknowledgements} 
We gratefully acknowledge valuable discussions with Drs. G. Olofsson, 
A. Heras, T. Prusti and G. Pilbratt. Some of the data presented here have been 
taken using ALFOSC, which is owned by the Instituto de Astrofisica de 
Andalucia (IAA) and operated at the Nordic Optical Telescope under 
agreement between IAA and the NBIfAFG of the Astronomical Observatory 
of Copenhagen. The comments and positve criticism of the referee, Dr. 
Bo Reipurth, which led to a more stringent paper, are gratefully acknowledged.
\end{acknowledgements}
%
%
%
%
%
%
%
%
%
%
%
%
%
%
%
%
%
%
%
%
%
%
%
%
%
%
%
%
%
%
%
%
%
%
%
%
\bibliographystyle{aa}
\bibliography{mf2}

\end{document}